\documentclass[superscriptaddress,groupedaddress,nofootnoteinbib,12pt]{article}  

\usepackage{amsmath}
\usepackage{amsfonts}
\usepackage{authblk}
\usepackage{bbm}
\usepackage{color}
\usepackage[dvipsnames]{xcolor}
\usepackage{graphicx,graphics}
\usepackage{epstopdf}
\usepackage{caption}
\usepackage{subcaption}
\usepackage{float} 
\usepackage{pifont}
\usepackage{titlesec}
\usepackage{etoolbox}
\usepackage{indentfirst}
\usepackage{jcappub}
\usepackage{braket}
\usepackage{ulem}
\usepackage{dsfont}
\usepackage{slashed}
\usepackage{tikz}
\usepackage{cancel}

\usepackage{upgreek}

\usepackage{bookmark,hyperref}
\hypersetup{
	colorlinks,
	citecolor=blue,
	filecolor=blue,
	linkcolor=blue,
	urlcolor=blue
}

\def\bb{\begin{equation}}
\def\ee{\end{equation}}
\def\ba{\begin{array}}
\def\ea{\end{array}}
\def\babc{\begin{subequations}}
\def\eabc{\end{subequations}}

\def\5{\hspace*{5mm}}
\def\2{{\scriptstyle\frac12}}

\def\M{\bar M}
\def\s{\sigma}
\def\g{\bar g}

\newcommand\redsout{\bgroup\markoverwith{\textcolor{red}{\rule[0.5ex]{2pt}{0.4pt}}}\ULon}

\tikzset{every node/.style={align=center}}

\begin{document}


\vspace{0.5in}

\begin{center}

\Large{\bf Inflation with the Trace Anomaly Action 

and Primordial Black Holes}

\vspace{0.5in}

\large{Gregory Gabadadze\,$^{a,b}$, David N. Spergel\,$^{b,c}$,  and Giorgi Tukhashvili$^d$}

\vspace{0.2in}

\large{\it $^a$Center for Cosmology and Particle Physics, Department of Physics, New York University, 726 Broadway, New York, NY, 10003, USA \\
\it $^b$ The Simons Foundation, 160 5th venue, New York, NY, 10010, USA \\
\it $^c$ Department of Astrophysical Sciences, Princeton University, Princeton, NJ, 08544, USA \\
\it $^d$Department of Physics, Princeton University, Princeton, NJ, 08544, USA }

\vspace{0.3in}

\end{center}

We study inflation in a recently proposed gravitational effective field theory 
describing the trace anomaly. The theory requires an additional scalar which is 
massless in the early universe. This scalar -- referenced as an anomalyon -- couples to the 
familiar matter  and radiation through the gauge 
field trace anomaly.  We derive a class of cosmological solutions that deviate 
from the standard inflationary ones only slightly, in spite of the fact that the  
anomalyon has a sizable time dependent background.  On the other hand, the scalar cosmological perturbations 
in this theory are different from the conventional inflationary perturbations. The inflaton and anomalyon perturbations mix, 
and one of the diagonal combinations  
gives the standard nearly scale-invariant adiabatic spectrum,  while the other combination 
has a blue power spectrum at short distance scales.  We argue that this blue spectrum 
can lead to the formation of primordial black holes (PBHs) at distance scales much shorter
than the ones tested in CMB observations. The resulting PBHs  can be heavy enough 
to survive to the present day universe. For natural values of the parameters involved 
the PBHs would constitute only a tiny fraction of the dark matter,  but  
with fine-tunings perhaps all of dark matter could be accounted by them.  
We also show that the theory predicts primordial gravitational waves which are 
almost identical to the standard inflationary ones.

\vspace{2in}


\newpage


\section{Introduction and Summary}\label{into_sum}


Quantum anomalies in field theories are quantum violations of symmetries present at the classical level,
even if these  symmetries are only approximate to begin with.  Consequences of quantum anomalies 
have long been explored. The well know examples are the decays of certain neutral pseudo-scalar mesons, 
such as a pion,  into two photons due to the axial anomaly \cite{Adler:1969gk,Bell:1969ts}, or the connection of the Hawking radiation \cite{Hawking:1974rv} to the gravitational trace anomaly \cite{Duff:1977ay,Capper:1975ig} (see, \cite {Christensen:1977jc}, and references citing it.).

 It is often convenient to have a local effective action that captures the respective quantum anomaly. The
 equations of motion obtained from such an action would contain the respective quantum information. 
 The Wess-Zumino-Witten term in the effective chiral Lagrangian of QCD is an example of such an action 
 for the axial anomaly \cite{Wess:1971yu,Witten:1983tw}. The 2D Polyakov action \cite{Polyakov:1981rd} 
 is  an example of such an action for a 2D trace anomaly. The analogous 4D action for 
 the 4D trace anomaly in dynamical gravitational field 
 would in particular be useful as it would capture quantum effects of spin-0,1/2, and 1 fields, and enable one to study them in terms of partial differential equations, 
 as opposed to the 
 cumbersome  correlation functions in a gravitational background \footnote{The spin-0,1/2,1, 
 fields are quantized in the presence of a classical dynamical gravitational field. This is a good approximation
 for physics below the Planck energy scale,  and we adopt it in this work. Gravity could also be quantized as an effective field theory, and its effects easily included in the formalism below, but this is not necessary for our goals.}. 
 
 The goal of this paper is to explore cosmological 
 consequences of such an effective action, in particular to study inflation 
 caused by a conventional single inflaton field. We will also study more general 
 Friedmann-Lemaitre-Robertson-Walker (FLRW) cosmological solutions.

 For reasons summarized below, there has to be at least two scalar fields in 
 the effective action during inflation. One of them is  the conventional inflaton of a single-field inflationary model, 
 and another one is a massless scalar related to the trace 
 anomaly. The questions we would like to address is whether such 
 a theory can retain the successful predictions of the inflationary paradigm, followed by an acceptable 
 FLRW phase, and whether the theory could  supply a new mechanism for the formation of primordial black holes (PBHs) 
 due to the quantum fluctuations  of the additional scalar related to the trace anomaly.

 The  local diffeomorphism invariant effective action for the trace anomaly that is weakly coupled
 below a certain  high energy scale $\M$ has the following form \cite{Gabadadze:2023quw}
\bb
S_{tot} = S(g)+ {\bar S}(\g)+ {\bar S}_{A}(\g,\sigma)\,,
\label{actiontotal}
\ee
with the two metrics related as follows,  $\g_{\mu\nu} = e^{-2\s} g_{\mu\nu}$.  Furthermore,     
\bb
S(g) = M^2 \int d^4x \sqrt{g} R(g)+\cdots\,,
\label{GR}
\ee
where the dots represent higher dimensional curvature invariants written in terms of $g$ and its derivatives, which  
are suppressed by the Planck mass  $M_{Pl}=\sqrt{2} M$.  Moreover, 
 \bb
{\bar S}(\g) = - \M^2 \int d^4x \sqrt{\g} R(\g)+\cdots\,,
\label{GRbar}
\ee
where the dots denote  curvature invariants 
written in terms of $\g$ and its derivatives, which are suppressed by the new high energy 
scale, $\M >> TeV$ (the latter is still below the Planck mass, $\M <<M_{Pl}=\sqrt{2} M$). 

Note that the sign of the Einstein-Hilbert (EH) term in (\ref {GRbar})  has to be opposite to the sign 
of the conventional EH term in (\ref {GR}) in order for the conformal scalar hidden in (\ref {GRbar})  
to have a right-sign kinetic  term. 

Last but not least, the anomaly  action ${\bar S}_{A}(\g,\sigma)$ of \cite{Riegert:1984kt,Fradkin:1983tg} equals to:
\bb
- \gamma^2 \int d^4 x\sqrt{\g} \left (  \sigma {\bar E} - 4 {\bar G}^{\mu \nu} {\bar \nabla}_\mu \sigma {\bar \nabla}_\nu \sigma 
	- 4 \left( {\bar \nabla} \sigma \right)^2 {\bar \nabla}^2 \sigma - 2 \left( {\bar \nabla} \sigma \right)^4   - \frac{\mathfrak{c}}{\gamma^2} \sigma {\bar W}_{\mu \nu \alpha \beta} {\bar W}^{\mu \nu \alpha \beta} \right),
\label{Riegertbar}
\ee
where ${\bar E}$ denotes  the Gauss-Bonnet invariant, ${\bar R}_{\mu \nu \rho \sigma} {\bar R}^{\mu \nu \rho \sigma} - 
4 {\bar R}_{\mu \nu} {\bar R}^{\mu \nu} + {\bar R}^2$, while 
${\bar G}_{\mu \nu}$ and ${\bar W}_{\mu \nu \alpha \beta}$  are the Einstein and Weyl tensors, respectively.
The coefficients $\gamma$ and $\mathfrak{c}$ depend on the number of particles of spin-0,1/2,1 contributing to the trace 
anomaly.\footnote{$\gamma$ and $\mathfrak{c}$ would also receive contributions 
from the gravitational sector if it were to be quantized.} 

The effective action of those massless  fields which define the values of $\gamma$ and $\mathfrak{c}$ should in general be included in 
(\ref {actiontotal}). These fields might have their own classical backgrounds, and/or could have significant perturbations in a background of other fields, depending on a particular cosmological scenario. In our case, we will introduce only an inflaton action in (\ref {actiontotal}) in the next section, 
but will ignore the action of all the massless fields that define $\gamma$ and $\mathfrak{c}$; this is a good approximation as long as the number of such fields is not unnaturally large. 

The  form of the action (\ref {actiontotal}) is 
set uniquely by the symmetries and anomalies of the theory, up to the cosmological 
terms for the metrics $g$ and  $\g$, which are put to zero. However, there are ambiguities 
of coupling the gravitational field to matter that we need to specify. 
Indeed, conventional matter and radiation fields could couple to the metric $g$, or
to a linear combination of $g$ and $\g$, with only a small admixture of $\g$ 
\cite {Gabadadze:2023quw}. Furthermore, the field $\s$ could be responsible for reproducing 
only the gravitational trace anomaly, as in  
\cite {Gabadadze:2023quw}, or alternatively, the same field $\s$ could be responsible for the gauge trace anomalies too. In the latter case $\s$ would have to couple to the gauge 
trace anomalies in the full quantum effective action \cite {Riegert:1984kt}. 
In the present work, we will choose the conventional matter and radiation to couple to the metric $g$, with no additional couplings  to $\g$ or $\s$, in the classical theory.  Furthermore, we will assume that the same field $\s$ that is responsible for the gravitational trace anomaly is also responsible for capturing the gauge trace anomalies;  therefore $\s$ will have couplings to the 
gauge trace anomalies in the full quantum effective action \cite {Riegert:1984kt}, but no other couplings to matter or radiation. Note that this is different from dilaton couplings. For that reason we will call $\s$ an {\it Anomalyon}. 

An anomalyon is massless during inflation, but would acquire a finite-temperature mass in the radiation dominated universe. More importantly, it will acquire a non-perturbatively 
generated potential and mass because of coupling to the QCD trace anomaly, or to the 
trace anomaly of a hypothetical high energy QCD-like theory, 
upon the respective phase transition to its confining phase. Such a mass 
would enables it to evade the fifth force constraints in the present-day 
universe, as it will be discussed in a separate work  \cite {followUpPaper}. 
Here we focus on the earlier, inflationary  and radiation dominated epochs 
of the universe, deferring the discussion of the high energy QCD-like, 
or QCD phases to a separate work \cite {followUpPaper}.

Before we turn to inflation there are a few important comments to be made 
about the 4D local diff-invariant weakly-coupled gravitational effective action for the trace anomaly (\ref {actiontotal}). These comments are of historical 
 nature and are gathered in this and the next paragraph.   
 The total effective action (\ref {actiontotal}) contains as its key ingredient the anomalous part (\ref {Riegertbar}) that 
 was first derived by Riegert \cite {Riegert:1984kt}, and simultaneously  by Efim Fradkin and Tseytlin \cite{Fradkin:1983tg} in a somewhat limited context, but was 
 placed on a solid effective field theory footing  by Komargodski and Schwimmer \cite{Komargodski:2011vj}. The very same action (\ref {Riegertbar}) was derived as a $SO(2,4)/ISO(1,3)$ 
 coset Wess-Zumino term in \cite{Gabadadze:2020tvt}, enhancing its standing as an effective action. This part of the effective action is often referenced as the Rieget action in the literature  \footnote{
 Riegert's work \cite{Riegert:1984kt} also proposed a {\it non-local} 
 effective action for the trace anomaly, which is sometimes referenced as 
 the {\it non-local} Riegert action, but it has a ghost since it can be rewritten as a local action with a massless scalar that has a four-derivative kinetic term \cite {Riegert:1984kt}. This latter action is not equivalent to the 
 local Riegert action considered in the present work, and it will not be used here.  See comments in \cite{Gabadadze:2023quw}.}.
 
 However, it was pointed out in \cite{Gabadadze:2023quw}, that if the local 
 and diff-invariant Riegert action (\ref {Riegertbar})  is 
 added to the action of General Relativity (GR) (\ref {GR}), one gets an 
 inconsistent theory. One way to see this
 inconsistency is to look  at nonlinear tree-level scattering 
 amplitudes; some of these amplitudes blow up in this theory,  with the blow up taking place at  arbitrarily low energies (this is similar to the discussions in \cite{Bonifacio:2020vbk} presented in 
 a different context). 
 To cure this inconsistency  Ref. \cite{Gabadadze:2023quw} proposed to amend the GR action by  the action 
 (\ref {GRbar}); this amendment preserves the symmetries and anomalies of the theory intact. 
 If the Riegert action is added to the the sum of (\ref {GR}) and (\ref {GRbar}),
 then the total action (\ref {actiontotal}) avoids the problem of the infinite strong coupling.  
 Indeed, the effective field theory 
 (\ref {actiontotal}) is weakly coupled at low energies, and becomes strongly coupled only at a certain high-energy scale  $\M$. While the value of  $\bar M$ is not known at present, it is presumed to be significantly higher than any Standard Model scale, 
 but should be much below the Planck scale. 
There are a finite number of higher dimensional terms suppressed by the scale $\M$  explicitly written 
in the Riegert action (\ref {Riegertbar}), but there is an infinite number of them  in (\ref {GRbar}) denoted by the dots. 
Interestingly, the entire theory (\ref {actiontotal}) can be embedded into the holographic
framework, providing a weakly-coupled high energy completion above the 
scale of $\M$ \cite{Gabadadze:2023tgi}. \footnote{The action (\ref {actiontotal}) without the 
Weyl squared term  but with a cosmological term for $\g$ was first proposed in Ref. \cite {Fernandes:2021dsb} in an entirely classical context without any reference to quantum 
trace anomaly, as an action that keeps the second order scale invariant equations of motion.
Some interesting classical solutions of these actions were found in \cite {Fernandes:2021dsb}, 
and \cite {Tsujikawa:2023egy}. It also emerged more recently in a different approach related to the trace anomaly, \cite{Karateev:2023mrb}.}

In this work we will consider the effective field theory (EFT)  (\ref {actiontotal}) in cosmological setting. We will find a class of 
FLRW cosmological solutions for the metric $g_{\mu\nu}$, such that  $\g_{\mu\nu}$ is proportional to  the Minkowski metric $\eta_{\mu\nu}$.
As a consequence, all curvature invariants written in terms  of $\g_{\mu\nu}$ in (\ref {GRbar}) are zero 
on these solutions.  As long as these solutions  have curvature invariants of $g_{\mu\nu}$  significantly smaller
than $M\sim M_{Pl}$, they are consistent solutions  of the EFT (\ref {actiontotal}).  On the other hand, 
the EFT  will  lead to strong coupling  of fluctuations on these cosmological solutions at a certain  
high-energy scale $\Lambda_c$. The latter can depend on the cosmological background and might in general be different from $\bar M$; 
the expression for the co-moving  scale $\Lambda_c$ is worked out in Appendix B. In this calculation we ignored the perturbations of all the massless fields that define the values of $\gamma$ and 
$\mathfrak{c}$. Moreover, fractional powers of  these parameters enter the definition of the strong scale since they multiply the nonlinear terms. For simplicity, we will not include the fractional powers of $\gamma$ and $\mathfrak{c}$ in the expression for the strong scale. All the above  are acceptable approximations as long as the values of $\gamma$ and $\mathfrak{c}$ are $\leq 100 $, or so.

Equipped with this EFT, we will consider an inflationary scenario by coupling (\ref {actiontotal}) 
to a single  inflaton in a conventional manner via the metric $g$. 

In section 2, we find exact cosmological solutions of the theory. Although the Friedmann equation gets modified 
by the new terms in (\ref {GRbar})  and (\ref {Riegertbar}), we will consider a regime well within the EFT where our solutions 
deviate very little from the conventional inflationary solutions. 

In section 3 we investigate  scalar perturbations above these solutions.
The novelty is that there are two scalar field perturbations, one for an inflaton and another one for an anomalyon.
The two mix, and the mixing can be diagonalized in a slow-roll approximation,  leading to the  separation of 
a ``mostly-inflaton" from ``mostly-anomalyon" perturbation. 
The ``mostly-inflaton"  perturbation provides the standard nearly scale-invariant spectrum of adiabatic perturbations. 
The ``mostly-anomalyon" perturbations have peculiarities:  for large distance scales they are  scale invariant and negligible
as compared to the ``mostly-inflaton" perturbations, but for short enough scales they have blue power-spectrum, and can dominate over the ``mostly-inflaton" perturbations. 

In Section 4 we show how this blue spectrum can lead to the formation of PBHs 
at the length scales much shorter than those best tested in CMB.  Some of these PBHs are 
heavy enough to survive to the present day universe. For natural values of the parameters 
of the theory the PBHs would constitute only a tiny fraction of dark matter in the universe,  
but with some tuning of the parameters they could describe the entire dark matter.  

In Section 5 we calculate the tensor perturbations and show that they approximate very closely  the conventional inflationary tensor perturbations.


\vspace{0.2in}


\section{Background Dynamics}\label{sect_1}


We will work with the theory (\ref {actiontotal}) described in the previous section,  but will write 
it in terms of the the metric $g_{\mu\nu}$ and $\s$ field only, using the expression 
for ${\g}_{\mu\nu} = e^{-2\s}g_{\mu\nu}$. Moreover, we will introduce a single 
inflaton field $\omega$, with its potential $V(\omega)$:
\begin{align}\label{Lag_beg}
	\nonumber \mathcal{L} = \sqrt{g} \Bigg\{ & M^2 \Big( 1 - \lambda^2 e^{-2 \sigma} \Big) R 
	- 6 M^2 \lambda^2 e^{-2 \sigma} \left( \partial \sigma \right)^2 \\
	{} & - \gamma^2 \Big[ \sigma E + 4 G^{\mu \nu} \partial_\mu \sigma \partial_\nu \sigma 
	- 4 \left( \partial \sigma \right)^2 \nabla^2 \sigma + 2 \left( \partial \sigma \right)^4 \Big]  + \mathfrak{c}~ \sigma W_{\mu \nu \alpha \beta} W^{\mu \nu \alpha \beta} \\
	\nonumber {} & - \frac12 \left( \partial \omega \right)^2 - V (\omega) \Bigg\} .
\end{align}
Here $\lambda={\M/M}$, and $E$ is the Gauss-Bonnet invariant,
\bb
E = R_{\mu \nu \rho \sigma} R^{\mu \nu \rho \sigma} - 4 R_{\mu \nu} R^{\mu \nu} + R^2\,,
\ee 
$G_{\mu \nu}$ is the Einstein  tensor,  and $W_{\mu \nu \alpha \beta}$ is the Weyl tensor, all written 
in terms of $g$. Because we rewrote the Riegert action (\ref {Riegertbar})  in terms of $g$, the signs of the 
second and fourth coefficients in the second line in (\ref {Lag_beg}) are different from those in (\ref{Riegertbar}).   
In the decoupling limit, $M\to \infty, \lambda \to 0, M\lambda = \M=const$, gravity decouples, and 
the anomalyon turns into   a Nambu-Goldstone boson of spontaneously broken conformal  invariance, with $\M$ 
being the symmetry breaking scale \cite{Gabadadze:2020tvt} (the anomalyon retains in this limit 
its couplings to the gauge trace anomalies, not shown in (\ref{Lag_beg})).

The sign in front of $\gamma^2$ is chosen in a way that in the decoupling limit $M \rightarrow \infty$, the forward scattering amplitude of $2 \sigma \rightarrow 2 \sigma$ is positive \cite{Komargodski:2011vj}. 

The variation of the Lagrangian w.r.t. $g_{\mu \nu}$, $\sigma$ and $\omega$ leads to the following equations respectively:
\bb\label{metric_var}
M^2 G_{\mu \nu} - M^2 \lambda^2 e^{-2 \sigma} \mathcal{J}_{\mu \nu} - 4 \gamma^2 \mathcal{K}_{\mu \nu} 
+ \mathfrak{c} ~\mathcal{C}_{\mu \nu} = \frac12 T_{\mu \nu} ,
\ee
\bb\label{sigma_var}
- 2 M^2 \lambda^2 e^{-2 \sigma} \left[ \mathcal{J} \right] - 8 \gamma^2 \left[ \mathcal{K} \right] - \gamma^2 E 
+ \mathfrak{c} ~W_{\mu \nu \alpha \beta} W^{\mu \nu \alpha \beta} = 0 ,
\ee
\bb\label{cov_omega_eq}
\nabla^2 \omega - \frac{d V (\omega)}{d \omega} = 0 .
\ee
$T_{\mu\nu}$ in our notations stands for the classical part of the stress-tensor of an inflaton. For the rest of the notations, please see Appendix \ref{app_EoM}. There exists a combination of the trace of (\ref{metric_var}) and (\ref{sigma_var}) in which $\sigma$ does not enter:
\bb\label{cov_anom_eq}
2 M^2 R - \gamma^2 E+ \mathfrak{c} ~W_{\mu \nu \alpha \beta} W^{\mu \nu \alpha \beta} 
= - \left[ T \right]\,.
\ee
This is the equation that captures the trace  anomaly \cite{Capper:1975ig,Duff:1977ay}. \\

For the  FLRW metric 
\bb\label{FRW_ansatz}
ds^2 = g_{\mu \nu}^{FRW} dx^\mu dx^\nu = - dt^2 + a^2 d \vec{x}^2\,,
\ee
the equations simplify. A convenient set of the resulting equations is as follows:
\bb\label{friedmann_like}
H^2 - \lambda^2 e^{-2 \sigma} \left( \dot{\sigma} - H \right)^2  + \frac{\gamma^2}{M^2}
\left[ \left( \dot{\sigma} - H \right)^4 - H^4 \right] = \frac{1}{6 M^2} T_{tt},
\ee
\bb\label{anom_eq_FRW}
\dot{H} + 2 H^2 - 2 \frac{\gamma^2}{M^2} H^2 \Big( \dot{H} + H^2 \Big) = \frac{1}{12 M^2} \Big( T_{tt} - 3 T_x^x \Big) ,
\ee
\bb\label{cont_eq}
\dot{T}_{tt} + 3 H \Big( T_{tt} + T_x^x \Big) = 0 .
\ee
The first equation is the $tt$ component of the Einstein equation for $g_{\mu \nu}$, the second is the anomalous equation (\ref{cov_anom_eq}), and the last one is eq. (\ref{cov_omega_eq}) which we rewrote as the continuity equation. ``$x$" denotes one of the spatial coordinates. All other equations can be obtained from these three, including the one for  $\sigma$ (see Appendix \ref{app_EoM}). Note that $\s=const.$ is not a solution of the above system, hence 
time evolution of $\s$ will be present in cosmology. The choice of the above  set of equations is convenient for the following reason. Multiplying the anomalous equation by $H$ and using the continuity equation to eliminate the pressure, we find that eq. (\ref{anom_eq_FRW}) reduces to:
\bb
\frac{1}{2 a^4} \partial_t \left[ a^4 \left( H^2 - \frac{\gamma^2}{M^2} H^4 - \frac{T_{tt}}{6 M^2} \right) \right] = 0\, .
\ee
The latter can be trivially integrated. Let us denote the integration constant by $\rho_c$,  and 
as customary, call it dark radiation, or Casimir energy  \cite{Fischetti:1979ue,Starobinsky:1980te,Hawking:2000bb}. The integration leads to significant simplification of the full set of cosmological equations:
\bb\label{friedmann_eq}
H^2 - \frac{\gamma^2}{M^2} H^4 = \frac{T_{tt}}{6 M^2} + \frac{\rho_c}{6 M^2 a^4}\, ,
\ee
\bb\label{sigma_H_eq}
\lambda^2 e^{-2 \sigma} \left( \dot{\sigma} - H \right)^2  - \frac{\gamma^2}{M^2} \left( \dot{\sigma} - H \right)^4
= \frac{\rho_c}{6 M^2 a^4} \,,
\ee
\bb
\dot{T}_{tt} + 3 H \Big( T_{tt} + T_x^x \Big) = 0 \,.
\ee
Equations (\ref{friedmann_eq}) and (\ref{sigma_H_eq}) show that the Hubble parameter is sourced by both the inflaton stress-tensor and by Casimir energy density, while the difference $\dot{\sigma} - H$ is sourced only by the Casimir energy density. 

Our objective is to study a (nearly) de Sitter solution supported by the inflaton potential. Since the presence of a sizable Casimir energy is incompatible with this, 
we set $\rho_c = 0$. In this case the equation (\ref{sigma_H_eq}) can be easily solved with
\bb
\sigma = \sigma_0 = \log \left( \frac{\sqrt{12} \lambda}{c} \,a \right)\,,
\ee
where $c$ is an integration constant. We choose the following parametrization 
\bb
c= \sqrt{12} \,\lambda \,a_g\,,
\label{c}
\ee
where $a_g$ denotes the value of the scale factor
when $\s$ vanishes. In a good approximation, $a_g$ is determined 
by the value of the scale factor  when the putative high-energy 
QCD-like theory  (or just QCD, if its high-energy counterpart is absent in the theory) 
undergoes a phase transition from the de-confining to confining phase and  
the anomalyon gets a non-perturbative potential and mass \cite{followUpPaper}.

In our conventions the present-day value of the scale factor is 1, implying 
that $a_g<<1$, and since $\lambda<<1$, we know tat $c<<<1$ (The quantity $cM_{Pl}$ gives rise to a correction to the Planck mass on the cosmological background, but this is negligible since
$c<<<1$.)  Equations (\ref{friedmann_eq}) and (\ref{sigma_H_eq}) are quadratic equations for $H^2$ and $\left( \dot{\sigma} - H \right)^2$, respectively; therefore there exist two branches of solutions for both equations. In this paper we consider only those solutions that are regular in the $\gamma \rightarrow 0$ limit. The complete classification of solutions of (\ref{friedmann_eq}) and (\ref{sigma_H_eq}) will be studied elsewhere.

\vspace{0.2in}


\section{Scalar Perturbations}

Let us now study small perturbations above the classical solution found in the previous section. For convenience, we 
switch from proper time $t$ to conformal time $\tau$ 
\bb
dt  = a ~d\tau\,,
\ee
and follow the notations of \cite{Riotto:2002yw}. The small perturbations 
of an inflaton and anomalyon above their backgrounds will be denoted by $\theta$ and $\pi$ respectively, and jointly with the graviton perturbations they will be defined as follows (in the subsequent  sections we set 
$M=1$, unless $M$ is shown explicitly):
\bb
g_{\mu \nu} = g_{\mu \nu}^{FRW} + a^2 h_{\mu \nu} + a^2 \bar{h}_{\mu \nu}\,,
\5\5\5
\omega = \omega_0 + \theta \,,
\5\5\5
\sigma = \sigma_0 + \pi .
\ee
Here, $h_{\mu \nu}$ is the transverse and traceless helicity-2 perturbation, while $\bar{h}_{\mu \nu}$ contains the scalar parts of the metric perturbation.




Following the standard procedure we decompose the scalar part of the metric perturbation in the following way:
\bb
\bar{h}_{00} = - 2 \phi \,,
\5\5\5
\bar{h}_{0 i} = \partial_i B \,,
\5\5\5
\bar{h}_{i j} = - \left( 2 \psi + \frac{1}{3} \pmb{\partial}^2 E \right) \delta_{i j} + \partial_i \partial_j E .
\ee
Under the gauge transformations,
\bb
x^\mu \rightarrow x^\mu + \xi^\mu ,
\5\5\5 \text{with} \5\5\5
\xi^\mu = (\xi^0 , \partial^i \beta) ,
\ee
the scalar perturbations transform as:
\bb
\phi \rightarrow \phi - (\xi^0)' - \mathcal{H} \xi^0 \,,
\5\5
\psi \rightarrow \psi + \mathcal{H} \xi^0 + \frac{1}{3} \pmb{\partial}^2 \beta \,,
\5\5
B \rightarrow B + \xi^0 - \beta' \,,
\ee
\bb
E \rightarrow E - 2 \beta \,,
\5\5\5
\theta \rightarrow \theta - \xi^0 (\omega_0)' \,,
\5\5\5
\pi \rightarrow \pi - \xi^0 (\sigma_0)' \,.
\ee
Then, Bardeen's gauge invariant variables read as follows:
\bb
\Phi = \phi + \frac{1}{a} \left[ a \left( B - \frac{E'}{2} \right) \right]' \,,
\5\5\5
\Psi = \psi - \frac{a'}{a} \left( B - \frac{E'}{2} \right) + \frac{1}{6} \pmb{\partial}^2 E \,,
\ee
\bb
\Theta = \theta + \left( B - \frac{E'}{2} \right) (\omega_0)' \,,
\5\5\5
\Pi = \pi + \left( B - \frac{E'}{2} \right) (\sigma_0)' .
\ee
The full Lagrangian of the scalar perturbations can be rewritten in terms of the Bardeen 
variables and is given by the following expression:
\begin{align}\label{scalar_perts_bardeen}
	\nonumber \mathcal{L}_S = & \frac{a^2}{2} \Big( T_{\tau \tau} + T_{xx} \Big) \Phi^2 - 6 a^2 \left( \Psi' + \mathcal{H} \Phi \right)^2 + 2 a^2 \left( \partial_i \Psi \right)^2
	+ 4 a^2 \Phi \partial_i^2 \Psi  \\
	\nonumber {} & + \frac{c^2}{12} \Big\{ 6 \left( \Psi' + \Pi' \right)^2 - 6 \left( \partial_i \Pi \right)^2 
	- 2 \left( \partial_i \Psi \right)^2
	- 4 \Phi \partial_i^2 \Psi 
	- 4 \Phi \partial_i^2 \Pi + 8 \Psi \partial_i^2 \Pi \Big\} \\
	{} & + 2 \gamma^2 \mathcal{H}^2 \left\{ 6 \left( \Psi' + \mathcal{H} \Phi \right)^2  - 2 \left( 1 - 2 \epsilon \right) \left( \partial_i \Psi \right)^2
	- 4 \Phi \partial_i^2 \Psi \right\} \\
	\nonumber {} & + \frac{4 \mathfrak{c}}{3} \sigma_0 \Big( \partial_i^2 \Psi + \partial_i^2 \Phi \Big)^2 \\
	\nonumber {} & + \frac{a^2}{2} \Big[ (\Theta ')^2 - \left( \partial_i \Theta \right)^2 - a^2 \frac{d^2 V}{d \omega^2_0} \Theta^2 \Big]
	+ a^2 \omega_0' \Theta \Big( \Phi' + 3 \Psi' \Big) - 2 a^4 \frac{d V}{d \omega_0} \Theta \Phi \,.
\end{align}
No approximations, besides that of the linearization in perturbations, have been made so far. In the above expression, $T_{xx}/a^2$ and $T_{\tau \tau}/a^2$ are the pressure and energy density induced by the inflaton. The first and the last lines in (\ref{scalar_perts_bardeen}) are the terms that appear in Einstein's GR, plus  a canonical scalar field \cite{Sasaki:1986hm,Mukhanov:1990me}, while the rest appear due to the $\bar{R}$, and the Riegert terms present in (\ref{Lag_beg}). Since the terms proportional to $\mathfrak{c}$ (not to be confused with $c$) would only contribute through fourth derivatives into the equation of motion, we expect them to become relevant at very high energies outside of the domain of our interest. 
Therefore, we disregard them. The constraint equation,
\bb
\Psi' + \mathcal{H} \Phi - \frac{c^2}{12 a^2} \left( \Psi' + \Pi' \right)
- \frac{2 \gamma^2}{a^2} \mathcal{H}^2 \Big( \Psi' + \mathcal{H} \Phi \Big)
= \frac{1}{4} \omega_0' \Theta ,
\ee
allows a significant simplification of (\ref{scalar_perts_bardeen}). To this end, let's introduce the ``physical" fields:
\bb
v = c \left( \Pi +\Psi \right) \,,
\5\5
u = a \Theta + z \Psi - \frac{c ~z}{12 a^2 \mathcal{Z}_0} v  \,,
\5\5 \text{with} \5\5
z \equiv a^2 \frac{\omega_0'}{a'} \,.
\ee
These field redefinitions together with the constraint equation can be used to eliminate
$\Phi$ and $\Psi$ from the action, which now becomes:
\begin{align}\label{scalar_perts_MS}
	\nonumber \mathcal{L}_S = & \frac12 \left( u' \right)^2 - \frac12 \left( \partial_i u \right)^2 
	+ \frac12 \frac{z''}{z} u^2 + \frac12 \left( 1 - \frac{c^2}{12 a^2 \mathcal{Z}_0} \right) \left( v' \right)^2 \\
	{} & - \frac12 \left( 1 - \frac{c^2}{12 a^2 \mathcal{Z}_0} - \frac{c^2 \gamma^2 \mathcal{H}^2 z^2 }{36 a^6 \mathcal{Z}_0^3} \right) \left( \partial_i v \right)^2 
	+ \frac{\mathcal{Z}_1^2}{2} v^2 + \mathcal{Z}_1 u v' + \mathcal{Z}_2 u v .
\end{align}
The first three terms in (\ref{scalar_perts_MS}) are the standard ones that appear in GR with 
a canonical scalar \cite{Sasaki:1986hm,Mukhanov:1990me}; the rest of the terms exist because 
the anomalyon. Once the background equations are used, $\mathcal{Z}_i$ can be expressed as:
\bb
\mathcal{Z}_0 = 1 - \frac{2 \gamma^2}{a^2} \mathcal{H}^2 \,,
\5\5\5
\mathcal{Z}_1 =  \frac{c \mathcal{H} z }{6 a^2 \mathcal{Z}_0} \left( 1 + \frac{\gamma^2 \mathcal{H}^2 z^2 }{2 a^4 \mathcal{Z}_0^2} \right) \,,
\ee
\bb
\mathcal{Z}_2 = \frac{c \mathcal{H}^2 z }{6 a^2 \mathcal{Z}_0} \left[ 1 + \epsilon - 2 \delta
+ \left( 1 - \frac{1}{\mathcal{Z}_0} \right) \epsilon \Big( 1 + \epsilon + 2 \delta - \eta \Big) 
- 4 \left( 1 - \frac{1}{\mathcal{Z}_0} \right)^2 \epsilon^2 \right]\,.
\ee
The slow roll parameters are defined as:
\bb
\epsilon \equiv - \frac{\dot{H}}{H^2} = 1 - \frac{\mathcal{H}'}{\mathcal{H}^2} \,,
\5\5\5
\delta \equiv 1 - \frac{\omega_0''}{\mathcal{H} \omega_0'} \,,
\5\5\5
\eta \equiv \frac{ \epsilon' }{\mathcal{H} \epsilon} \,.
\ee
First, note that when $\gamma \neq 0$, the velocity of a particle described by $v$ is less than 1. 
This is due to  the fact that the sign of $\gamma^2$ was chosen for superluminalities to be absent from the scalar sector. 
Note that the mixings between the two scalar perturbations in (\ref{scalar_perts_MS}) are proportional to both $z \propto \epsilon^{1/2}$ and $c$. During inflation $\epsilon \ll 1$, and 
we'll see in the next section that the parameter $c$ is tiny, so the perturbations evolve essentially independently from one another during inflation, radiation, and matter dominated epochs. 
Since we are dealing with a spacetime that approximates  dS, we can neglect the terms that are $\mathcal{O} \left( c/ a \right) $ (as a reminder $\mathcal{H} / a = H \approx const$). Using the above approximations, the Lagrangian (\ref{scalar_perts_MS}) can be rewritten as:
\bb\label{lagrangian_u_v_simp}
\mathcal{L}_S = \frac12 \left( u' \right)^2 - \frac12 \left( \partial_i u \right)^2 
+ \frac12 \frac{z''}{z} u^2
+ \frac12 \left( v' \right)^2
- \frac12 \left( \partial_i v \right)^2  + \frac{\mathcal{Z}_1^2}{2} v^2 + \mathcal{Z}_1 u v' + \mathcal{Z}_2 u v\,.
\ee
Furthermore, we have:
\bb
\mathcal{Z}_0 \approx 1\,,
\5
\mathcal{Z}_1 \approx \frac{c H \sqrt{\epsilon}}{3} \,,
\5
\mathcal{Z}_2 \approx - \frac{ \mathcal{Z}_1 }{\tau} \,
\5
z^2 \approx 4 a^2 \epsilon \,
\5
\frac{z''}{z} \approx \frac{1}{\tau^2} \Big( 2 + 6 \epsilon - 3 \delta \Big) .
\ee
Since $\mathcal{Z}_1' = \mathcal{O} \left( \epsilon^{3/2} \right)$, we can regard it as effectively constant. Moreover,  there exist field redefinitions that allow approximate diagonalization of the Lagrangian (\ref{lagrangian_u_v_simp}) (only the terms that are up to linear in $\epsilon$ are kept):
\bb
u = \left( 1 - \frac{\tau^2 \mathcal{Z}_1^2}{8} \right) \mathfrak{u} + \frac{\tau \mathcal{Z}_1}{2} \upupsilon \,,
\5\5\5
v = \left( 1 - \frac{\tau^2 \mathcal{Z}_1^2}{8} \right) \upupsilon - \frac{\tau \mathcal{Z}_1}{2} \mathfrak{u} .
\ee
In terms of these variables the Lagrangian (\ref{lagrangian_u_v_simp}) becomes:
\bb
\mathcal{L}_S =
\frac12 \left( \mathfrak{u}' \right)^2 - \frac12 \left( \partial_i \mathfrak{u} \right)^2 
+ \frac12 \left( \frac{z''}{z} + \frac{ \mathcal{Z}_1^2 }{4} \right) \mathfrak{u}^2 
+ \frac12 \left( \upupsilon' \right)^2 - \frac12 \left( \partial_i \upupsilon \right)^2 
+ \frac{ \mathcal{Z}_1^2 }{8} \upupsilon^2
+ \mathcal{O} \left( \epsilon^{3/2} \right)\,.
\ee
Equations of motion that follow from the above Lagrangian in Fourier space read:
\bb
\mathfrak{u_{\pmb{k}}}'' + \left( k^2 - \frac{z''}{z} - \frac{ \mathcal{Z}_1^2 }{4} \right) \mathfrak{u_{\pmb{k}}} = 0 \,,
\5\5\5
\upupsilon_{\pmb{k}}'' + \left( k^2 - \frac{ \mathcal{Z}_1^2 }{4} \right) \upupsilon_{\pmb{k}} =0\,.
\ee
On sub-Hubble scales, $k \gg a H $ or $ - k \tau \gg 1 $, we impose the following quantization conditions:
\bb\label{u_quantization}
\mathfrak{u}_{\pmb{k}}^\star \mathfrak{u}_{\pmb{k}}' - \mathfrak{u}_{\pmb{k}} (\mathfrak{u}_{\pmb{k}}^\star)' = - i \,,
\5\5\5
\upupsilon_{\pmb{k}}^\star \upupsilon_{\pmb{k}}' - \upupsilon_{\pmb{k}} (\upupsilon_{\pmb{k}}^\star)' = - i .
\ee
Solutions for $\mathfrak{u}$ and $\upupsilon$ that satisfy the quantization conditions 
when $- k \tau \gg 1$ and $k \gg \mathcal{Z}_1 / 2$ are:
\bb\label{sol_anom_fluc}
\upupsilon_{\pmb{k}} = \frac{1}{\sqrt{2}} \left( k^2 - \frac{ \mathcal{Z}_1^2 }{4} \right)^{-1/4} 
\exp \left( - i \tau~ \sqrt{k^2 - \frac{ \mathcal{Z}_1^2 }{4}} \right) ,
\ee
\bb
\mathfrak{u}_{\pmb{k}} = \frac{\sqrt{\pi}}{2} e^{i \left( \nu + \frac12 \right) \frac{\pi}{2} } \sqrt{- \tau} 
~H^{(1)}_\nu \left( - \tau~ \sqrt{k^2 - \frac{ \mathcal{Z}_1^2 }{4}} \right) ,
\ee
with
\bb
\nu = \frac{3}{2} + 2 \epsilon - \delta + \mathcal{O} \Big( \epsilon^2, \delta^2 \Big) .
\ee
The three-dimensional Ricci curvature,
\bb
{}^{(3)} R = \frac{4}{a^2} \pmb{\partial}^2 \psi  + \frac{2}{3 a^2} \pmb{\partial}^2 \pmb{\partial}^2 E
= \frac{4}{a^2} \pmb{\partial}^2 \left( \psi + \frac{1}{6} \pmb{\partial}^2 E \right)  ,
\ee
can be used to build the following gauge invariant combination:
\bb
\psi + \frac{1}{6} \pmb{\partial}^2 E + \frac{a'}{a~ \omega_0'} \theta = \Psi + \frac{a'}{a~ \omega_0'} \Theta
= \frac{u}{z} + \frac{c}{12 a^2 \mathcal{Z}_0 } v
\equiv \mathcal{R}^\Theta .
\ee
Due to the existence of the anomalyon field $\sigma$, there is another gauge invariant combination (built in an analogy with $\mathcal{R}^\Theta$):
\bb
\psi + \frac{1}{6} \pmb{\partial}^2 E + \frac{a'}{a~ \sigma_0'} \pi = \Psi + \frac{a'}{a~ \sigma_0'} \Pi
= \Psi + \Pi = \frac{v}{c}
\equiv \mathcal{R}^\Pi .
\ee
The curvature perturbations in terms of $\mathfrak{u}$ and $\upupsilon$ are:
\bb
\mathcal{R}^\Theta_{\pmb{k}} \approx 
\frac{1}{2 a \sqrt{\epsilon}} \left( 1 + \frac{c^2 \epsilon}{72 a^2} \right) \mathfrak{u}_{\pmb{k}}
- \frac{c^3 \epsilon}{864 a^4} \upupsilon_{\pmb{k}} ,
\ee
and
\bb
\mathcal{R}^\Pi_{\pmb{k}} \approx 
\frac{1}{c} \left( 1 - \frac{c^2 \epsilon}{72 a^2} \right) \upupsilon_{\pmb{k}}
+ \frac{\sqrt{\epsilon}}{6 a} \mathfrak{u}_{\pmb{k}} .
\ee
When $k^2 \gg \mathcal{Z}_1^2 / 4$ and $-\tau k \ll 1$, the respective power spectra read:
\bb\label{PS_theta}
\mathcal{P}_{\mathcal{R}^\Theta} = \frac{k^3}{2 \pi^2} \left| \mathcal{R}_{\pmb{k}}^\Theta \right|^2 =
\frac{1}{4 \epsilon } \left( \frac{H}{2 \pi} \right)^2  
\left( \frac{k}{a H} \right)^{3 - 2 \nu} \left( 1 + \frac{c^2 \epsilon}{36 a^2} \right) \,,
\ee
\bb \label{PS_Pi}
\mathcal{P}_{\mathcal{R}^\Pi} = \frac{k^3}{2 \pi^2} \left| \mathcal{R}_{\pmb{k}}^\Pi \right|^2 =
\frac{k^2}{4 c^2 \pi^2} + \frac{H^2 \epsilon}{144 \pi^2} .
\ee
The spectral index, $n_{\mathcal{R}^\Theta}$, for $\mathcal{P}_{\mathcal{R}^\Theta}$ can be expressed in terms of the slow roll parameters:
\bb
n_{\mathcal{R}^\Theta} - 1 \equiv 3 - 2 \nu =
2 \delta - 4 \epsilon .
\ee
Formula (\ref{PS_theta}) shows that $\mathcal{R}^\Theta_{\pmb{k}}$ coincides with the adiabatic perturbation up to a correction that is both small $\propto c \epsilon<<<1$, and decaying $\propto a^{-2}$. The last formula is a bit peculiar; it shows the existence of the blue power spectrum in $\mathcal{P}_{\mathcal{R}^\Pi} $ and a small, scale-invariant correction to it. The blue spectrum in $\mathcal{P}_{\mathcal{R}^\Pi} $ is somewhat unusual and we will give an explanation for it in the next section.  The second, scale-invariant term in $\mathcal{P}_{\mathcal{R}^\Pi} $ originates from the mixing between the $\sigma$ perturbation and the scalar parts of the metric perturbation.

We can build another gauge invariant curvature perturbation:
\bb
\zeta \equiv \psi + \frac{1}{6} \pmb{\partial}^2 E + \mathcal{H} \frac{ \delta T^\tau_\tau }{ \left( T^\tau_\tau \right)' }
 = \Psi - \frac{ \mathcal{H} }{ \left( T^\tau_\tau \right)' } \left[ \frac{1}{a^2} \omega_0' \Theta' + \frac{d V (\omega_0)}{d \omega_0} \Theta - \frac{1}{a^2} \Phi \left( \omega_0' \right)^2  \right] .
\ee
In addition we define gauge invariant comoving density contrasts:
\bb
\Delta^\Theta \equiv \frac{ \left( T^\tau_\tau \right)' }{ T^\tau_\tau }
\left( \frac{\theta}{\omega_0'} - \frac{ \delta T^\tau_\tau }{ \left( T^\tau_\tau \right)' } \right)
= \frac{ \left( T^\tau_\tau \right)' }{ \mathcal{H} T^\tau_\tau } \Big( \mathcal{R}^\Theta - \zeta \Big) \,,
\ee
\bb \label{PsOverdensity}
\Delta^\Pi \equiv \frac{ \left( T^\tau_\tau \right)' }{ T^\tau_\tau }
\left( \frac{\pi}{\sigma_0'} - \frac{ \delta T^\tau_\tau }{ \left( T^\tau_\tau \right)' } \right)
= \frac{ \left( T^\tau_\tau \right)' }{ \mathcal{H} T^\tau_\tau } \Big( \mathcal{R}^\Pi - \zeta \Big) .
\ee
The Poisson equation can be reduced to:
\bb
\mathcal{Z}_0 \partial_i^2 \Psi = \frac{a^2}{4} T^\tau_\tau \Delta^\Theta + \frac{c}{12 a^2} \partial_i^2 v 
+ \frac{c \mathcal{H}}{4 a^2} v' .
\ee
If we ignore terms $\propto c$ and keep in mind that $\mathcal{Z}_0 \approx 1$, we would recover the Poisson equation for  GR perturbations. On super-Horizon scales $\mathcal{R}^\Theta \approx \zeta$. If we keep this in mind,
\bb
\Delta^\Pi_{\pmb{k}} \approx
\frac{ \left( T^\tau_\tau \right)' }{ \mathcal{H} T^\tau_\tau } \Big( \mathcal{R}^\Pi_{\pmb{k}} - \mathcal{R}^\Theta_{\pmb{k}} \Big) .
\ee
The power spectrum of $\Delta^\Pi$ is proportional to the sum of power spectra of comoving curvature perturbations:
\bb\label{PS_Delta_PI}
\mathcal{P}_{\Delta^\Pi} \approx \left( \frac{ \left( T^\tau_\tau \right)' }{ \mathcal{H} T^\tau_\tau } \right)^2 
\Big( \mathcal{P}_{\mathcal{R}^\Theta} + \mathcal{P}_{\mathcal{R}^\Pi} \Big)
= \left( - 2 \epsilon ~\frac{1-2 \gamma^2 H^2}{1-\gamma^2 H^2 } \right)^2 \Big( \mathcal{P}_{\mathcal{R}^\Theta} + \mathcal{P}_{\mathcal{R}^\Pi} \Big).
\ee
It is clear that the blue spectrum could dominate over the (nearly) scale-invariant one without leaving EFT (see more below).


\section{Primordial Black Holes}

There is a tremendous amount of literature on the interesting subject of PBH formation,
see, e.g., \cite{Zeldovich:1967lct,Carr:1974nx,Carr:1975qj,Novikov1979,Khlopov:1985fch,Carr:1993aq,Ivanov:1994pa,Garcia-Bellido:1996mdl,Sasaki:2018dmp,Garcia-Bellido:2017mdw,Ballesteros:2017fsr,Carr:2019kxo,Carr:2023tpt,Carr:2020xqk,Mishra:2019pzq} (see also \cite{Khlopov:2008qy,Belotsky:2018wph,Heydari:2021gea,Ashrafzadeh:2024oll}), and reference therein.
Our goal in this section will be to outline a new mechanism for the PBH formation due to the anomalyon fluctuations. We will have nothing to add to the important question of the calculation of the abundance of the primordial PBHs \cite {Press:1973iz,Young:2014ana,Green:2004wb}. Since we are after the estimates of the abundance, we will follow the standard Press-Schechter framework \cite {Press:1973iz}, even though other elaborate formalisms 
could give more justified and precise results \cite{Young:2014ana,Green:2004wb}.

To reiterate, the power spectrum of the ``mostly-anomalyon" co-moving 
curvature perturbation calculated in the previous section 
is:
\bb\label{PS_Pi2}
\mathcal{P}_{\mathcal{R}^\Pi} \simeq
\frac{k^2}{2 c^2 \pi^2 M_{Pl}^2} + \frac{H^2 \epsilon}{72 \pi^2  M_{Pl}^2} .
\ee
The expression  (\ref {PS_Pi2})  has two distinct terms on the right hand side: 
the first one grows in the blue part of the spectrum and could be large, while  the second one is 
scale invariant but small. The blue part could give rise to novel features at short distance 
scales without affecting too much the large scale  predictions. 

More quantitatively,  
at low momenta corresponding to large distance scales, such as the CMB scale,  both terms 
in (\ref{PS_Pi2}) are sub-leading to the ``mostly-inflaton" nearly scale-invariant 
power spectrum 
\bb\label{PS_theta_approx}
\mathcal{P}_{\mathcal{R}^\Theta} \simeq \frac{1}{2 \epsilon } \left( \frac{H}{2 \pi M_{Pl} } \right)^2  
\left( \frac{k}{a H} \right)^{3 - 2 \nu}\,.
\ee
The latter is responsible for the  power  $ \mathcal{P}_{\mathcal{R}^\Theta} =2.1 \times 10^{-9}$ needed at the co-moving CMB pivot scale of $k_{CMB}=0.05 Mpc^{-1}$  \cite{WMAP:2003elm,WMAP:2003ivt}, as in conventional inflation.  
On the other hand, the $k^2$ dependent  term  
in the ``mostly-anomalyon" power (\ref {PS_Pi2}) dominates over the scale-independent term for 
the momenta  large  compared  to the CMB pivot scale; it  also dominates over   (\ref {PS_theta_approx}). 
In that regime, the blue power -- if sufficiently strong -- 
could lead to the the formation of PBHs.

Therefore, we focus on the regime where the blue part of the spectrum is dominant 
\bb\label{cond_1}
\frac{k^2}{2 c^2 \pi^2 M_{Pl}^2} \gg \frac{H^2 \epsilon}{72 \pi^2  M_{Pl}^2}\,,
\5\5\5 \rightarrow \5\5\5
k \gg \frac{c H \sqrt{\epsilon}}{6}\,.
\ee
Note that both $\epsilon<<1$, $c<<1$, and the above condition can be  achieved
within the effective field theory.  In this approximation the ``mostly-anomalyon" fluctuations 
reduce to those of a pure anomalyon. Note that the inequality in (\ref {cond_1}) does not preclude
the existence of the super-horizon modes.

A large co-moving wavenumber that is still within the effective field theory,  
and could lead to a large power spectrum is given by 
\bb
k_{PBH} \equiv q\,c M_{Pl}\,=q \sqrt{12} \,a_g \M\,,
\label{kpbh}
\ee 
where $q<<1$ is a parameter that guarantees that the value of $k_{PBH}$  is below  
the cutoff of the co-moving momenta, $cM_{Pl}$, and that the 
nonlinear terms are sub-leading to the linear ones. 
The power  spectrum for this high momentum mode is
\bb\label{PS_PBH}
\mathcal{P}^{PBH}_{\mathcal{R}^\Pi} \simeq
\frac{q^2}{2  \pi^2}.
\ee
The PBH formation  requires the power spectrum to be significantly larger than it is 
for CMB, albeit at a much shorter scale \cite{Garcia-Bellido:2017mdw,Ballesteros:2017fsr}. The value of the power spectrum that would be favorable for the PBH formation for a given $k$, 
depends on how the power spectrum scales with $k$, 
and could range in the interval, $10^{-4}-10^{-2}$. This sets the value of 
$q<<1$. We will find below that  in our case $\mathcal{P}_{\mathcal{R}} 
\sim 10^{-4} $, and $q\sim 5 \times 10^{-2}$.

In the ``blue" regime the properly normalized high momentum anomalyon 
mode has the wavefunction that depends on conformal time as follows:
\bb
\upupsilon_{\pmb{k}} = \frac{1}{\sqrt{2k}}
\exp \left( - i \tau k \right).
\label{wavef}
\ee
This mode oscillates at both  sub-horizon or super-horizon scales (its physical wave-number is redshifted with the expansion, and one can still talk about the 
sub - and super-horizon modes). The amplitude of a super-horizon mode  
is then frozen-in. Moreover, the 
curvature perturbations are caused by these super-horizon modes.  They  
will re-enter the Hubble volume sometime during the radiation domination. 
However, the anomalyon super-horizon  modes could 
carry high enough power at shorter scales and could collapse into PBH's upon their 
re-entry to the Hubble volume during the radiation dominated epoch \cite{Garcia-Bellido:2017mdw,Ballesteros:2017fsr}. 

The anomalyon perturbations (\ref{sol_anom_fluc}) are not classical, and unlike the inflaton perturbations they do not grow at superhorizon scales to become classical. However, they do not have to be classical to form black holes: quantum fluctuations carry energy, and this energy gravitates according to the equivalence principle. 
In more concrete terms, the power spectrum of a quantized field calculated in (\ref{PS_Pi}) 
is equated to the power spectrum of the density contrast (\ref{PsOverdensity}) 
which then is regarded as a classical averaged density contrast coupled to GR. This averaging 
is taking place when the perturbations re-enter the horizon (i.e., the averaging is done over a causally connected Hubble volume). The resulting expression is then used below in this section as a classical density contrast in the Press-Schechter formalism. Once this framework is adopted, the classical calculations for the collapse of over-densities apply. 
Indeed, the amplitude square of the density contrast that is re-entering the Hubble volume will be computed below to be of the order of $2.83\times 10^{-3}$. This density contrast needs to become of order one to collapse into a  PBH. 
One can crudely estimate the comoving time it takes for a density contrast amplitude to evolve non-linearly from $10^{-3/2}$  to $\sim 1$. Using a simple spherical collapse model with 
the Friedmann equation (see, e.g., \cite{Carr:1975qj}) one finds that the density contrast will grow to the threshold of the PBH collapse in a co-moving time proportional to $10^{3/2}/H_{re}$, where $H_{re}$ is the value of the Hubble parameter when the perturbation re-entered the Hubble volume.  Hence, the value of the Hubble parameter at the moment when the anomalyon overdensity is at the verge of collapsing will have to be about 10-100 times smaller than its value when the perturbation re-entered the Hubble volume (the uncertainty is due to the proportionality coefficient in the relation above).

Before we go further, We need to  explain the seemingly un-physical dependence 
of the power spectrum (\ref {PS_Pi2}) on the co-moving momentum $k$. 
To do so, we rewrite (\ref {PS_Pi2})  as follows:
\bb
\mathcal{P}_{\mathcal{R}^\Pi} \simeq
\frac{1}{48 \pi^2 {\bar M}^2} \,\frac{k^2}{a(t)^2}\,e^{2\sigma_0(t)} + \frac{H^2 \epsilon}{72 \pi^2  M_{Pl}^2} ,
\label{PS_phys}
\ee
which makes it clear that the power spectrum depends on a physical momentum $k/a$, and on the classical solution  for the anomalyon, $\s_0 = ln ( a(t)/a_g)$ in a way that  there is an exact cancellation of the scale factor. As a result, the power spectrum  (\ref {PS_Pi2}) does not get redshifted. 

The expression (\ref {PS_phys}) can be used as follows.
A perturbation exiting the Hubble volume during inflation will have its physical wave-number equal the value of the Hubble parameter at inflation, $H_{inf}$. Therefore, the power spectrum of such a perturbation will take the form:
\bb
\mathcal{P}_{\mathcal{R}^\Pi} \simeq  \frac{H_{inf}^2}{48 \pi^2 {\bar M}^2} \,\frac{a_{inf}^2}{a_g^2}\,,
\label{PS_aa} 
\ee
where $a_{inf}$ is the value of the scale factor at the moment when the given mode exits the inflationary Hubble volume. It is clear from the above that the value of $a_g$ should not be too large as compared to $a_{inf}$, otherwise the anomalyon power spectrum would be too small  to  be interesting for PBH formation.  For instance, if there are no other confining gauge fields above the QCD scale, them $a_g$ will be set by QCD, and the corresponding power spectrum in (\ref {PS_aa}) will be unobservable, regardless on the values of $H_{inf}$ and $\M$. 
In what follows we will be assuming that the anomalyon acquires a non-perturbative potential and mass at some high energy scale significantly above the standard model energy scale,  where  
$a_g$ gets fixed (for instance, one could consider a QCD-like gauge theory at some high energy scale, that undergoes a phase transition from the de-confining to confining phase soon after the inflation and reheating). 

As it is clear from (\ref {PS_aa}) the anomalyon modes that exit the inflationary Hubble 
volume by the end of inflation will carry the highest power. Some of these modes will have
$\mathcal{P}_{\mathcal{R}} \sim 1 $, and will lead to the formation of causally disconnected 
universes, similarly to the large inflaton fluctuations during the early states of 
chaotic inflation \cite{Linde:1983gd}. On the other hand, there will always be  modes that exit the Hubble volume a few e-folds before the end of inflation, which could have 
$\mathcal{P}_{\mathcal{R}} \sim 10^{-4} $, if the value of $a_g$ is chosen 
appropriate. Since there is significant flexibility in 
setting the value of $a_g$ by some unknown 
high energy gauge dynamics, we will assume that such 
modes can always be attained.

The mass of  a PBH is proportional to the mass stored in 
the Hubble volume at the time 
when the high-power anomalyon perturbation re-enters it
\bb
M_{PBH}= \alpha\, \frac{4\pi \rho}{3 H_{re}^3}\,,
\label{Mpbh}
\ee
 where $\rho = \pi^2 g_{*}(T) T^4/30$ is the energy density of the $g_{*}(T)$ relativistic degrees of freedom 
at the temperature $T$ in the radiation dominated universe.  $H_{re}$ is the value of the Hubble parameter 
at the time when the high-power perturbation is re-entering the Hubble volume, and the overall multiplier 
$\alpha$ is a constant that parametrizes the efficiency of a large amplitude perturbation to collapse into a 
PBH once it re-enters the horizon; it is estimated to be $\alpha =  (3)^{-3/2} \simeq 0.2$ \cite{Carr:1975qj}.  

Consider now a primordial  anomalyon perturbation that exited the inflationary Hubble volume with a physical 
momentum $k_{phys}^{exit} \simeq H_{inf}$, where $H_{inf}$ is approximated 
by a constant during the quasi-de Sitter inflationary expansion. Let us suppose the 
exit took place when there were
$n$ number of e-folds  remaining before the end of inflation, where $n$ can be 
any number satisfying $n\leq 60$; we denote the value of the scale factor at this point by 
$a_{inf}(n)$.  This perturbation will have its physical wavenumber red-shifted.  It will 
re-enter the Hubble volume during the radiation dominated epoch when its physical momentum 
$k_{phys}^{re}$ becomes equal to the value of the Hubble parameter at the moment of re-entry, $H_{re}$ 
\bb
k_{phys}^{re} = k_{phys}^{exit} 
\left (   {a_{inf}(n) \over a_{re} }\right ) 
= H_{inf} \left (   {a_{inf}(n) \over a_{re} }\right )\,=H_{re}\,.
\label{kred}
\ee
Here we denoted the value of the scale factor at the re-entry by $a_{re}$. 
The last equation in the above relations can be rewritten  as follows:
\bb
 {H_{inf}  \over T_{eq}}  {a_{inf}(n) \over a_{eq}}\,={\pi \sqrt{g_*(T)}\over 3 \sqrt{10}}\,{T \over M_{Pl}}\,,
\label{Hre}
\ee
where $T_{eq}$ and $a_{eq}$ are the CMB temperature and scale factor, respectively, 
at the matter-radiation equality, and $g_*(T)$ is the number of relativistic degrees 
of freedom at the temperature $T$. Both $a_{inf}(n)$ and $a_{eq}$ depend of an arbitrary normalization of the scale factor, 
however, the ratio $a_{inf}(n)/a_{eq}$ appearing in (\ref {Hre}) is independent of 
this arbitrariness, and depends on the value of $n$.

Therefore, different values  of $n$ will lead to the super-horizon 
perturbations re-entering a Hubble volume at different temperatures $T$, which 
can be calculated form (\ref {Hre}) .  

Having established this, we can now return to the 
mass of a PBH and express it in terms of $T$ or $k$: 
\begin{align}\label{MpbhT}
    \nonumber M_{PBH}\simeq & M_{\odot} \left( \frac{\alpha}{0.2} \right) \left({106.75 \over g_{*}(T_f)} \right)^{1/2} \left({100\,MeV \over T } \right)^2 \\
    \simeq & 5 \times 10^{-16} M_{\odot}
\left( \frac{\alpha}{0.2} \right) \left({106.75 \over g_{*}(T_f)} \right)^{1/6} \left( \frac{7 \times 10^{13} Mpc^{-1}}{k} \right)^2 \,,
\end{align}
where $M_{\odot} \simeq 2 \times 10^{33} g$ is the solar mass. The second equality is obtained by invoking the entropy conservation \cite{Inomata:2017okj}.

If a PBH is heavier than $10^{15}{g}$, it would not have had enough time to evaporate entirely due to  the Hawking radiation; 
for PBHs to be heavier than $10^{15}{g}$  they need to form after the temperature dropped below  $10^8\,GeV$. 
For instance, a PBH formed  right before the electroweak phase transition, at $T\sim 1TeV$, would have a mass 
$6 \times 10^{-8} M_{\odot}$, while the one formed before the QCD phase transition would be as heavy as 
$ 6 \times 10^{-2} M_{\odot}$. The formation of supermassive PBHs with mass $>> 10^3 M_{\odot}$, would 
potentially   be  in conflict with the CMB spectral distortion constraints \cite{Hooper:2023nnl}.
In our framework, there is a natural upper cutoff on the PBH masses simply because the mechanism
can only work for high co-moving momenta exiting the inflationary Hubble volume 
a few e-folds before the inflation.

Besides the above long-lived PBHs it is also interesting to consider lighter  PBHs which evaporate around 
the time of Big Bang Nucleosynthesis, $t\sim (0.1 -1)$ sec. These would need to have mass 
$M_{PBH}^{1sec}\sim 10^{-25} M_{\odot}$, and would require their formation temperature 
to be, $T\sim 10^{12}\,GeV$. 

To determine the PBH abundance, we adopt the Press-Schechter 
approach \cite{Press:1973iz} by assuming the Gaussian distribution function for over-densities smoothed
out at a comoving length scale $R = k^{-1}$. 
The variance of the Gaussian random field of interest, 
which in our case is the fluctuation of the  anomalyon field, can be calculated as follows:
  \bb
\Delta^2_{ {\Delta^\Pi}} =  \int _0^\infty {dq\over q} \,w^2 (qR)\,\mathcal{P}_{\Delta^\Pi} (q)\,,
\label{var}
\ee
where we should substitute the power spectrum (\ref {PS_Pi2}), and as customary  
adopt the Fourier transform of the volume-normalized Gaussian window function $w(x)= e^{-x^2/2}$ \cite{Young:2014ana}.  The result of this calculation is 
\begin{align}\label{varA}
    \Delta_{ {\Delta^\Pi}}^2 \simeq & 3.42 \times 10^{-102}
\left( \frac{\alpha}{0.2} \right) \left({106.75 \over g_{*}(T_f)} \right)^{1/6} \left( \frac{\epsilon_{rad}}{c} \right)^2  \frac{M_\odot}{M_{PBH}} \,,
\end{align}
where $\epsilon_{rad}=2$ is the epsilon parameter during the radiation dominated epoch.
Furthermore,  
the PBH abundance at the scale of their  formation is \cite{Green:2004wb}
\bb\label{abundance}
\Omega _{form} = 2 \int_{\Delta_{th}} ~d \Delta ~P(\Delta) 
\simeq Erfc \left( \frac{\Delta_{th}}{\sqrt{2} \Delta_{ {\Delta^\Pi}}} \right) ,
\ee
where $Erfc$ denotes the complementary error-function, $\Delta_{th} $ is the threshold value of the overdensity to form a black hole. In \cite{Carr:1975qj}, the latter is approximated by the speed of sound during the radiation dominated era, $\Delta_{th} \approx c_{so} = 1/3$, though a more recent numerical analysis shows $\Delta_{th} \approx 0.45 \pm 0.02$ \cite{Musco:2004ak} (see \cite{Harada:2013epa} for analytic evaluations of $\Delta_{th}$ that are closer to this numerical value).
Using (\ref {varA}) in (\ref {abundance}), and taking into consideration that $\Delta_{th} \gg \Delta_{ {\Delta^\Pi}}$, we get an approximate expression for the PBH abundance 
at their formation
\bb\label{abundanceA}
\Omega _{form} \simeq \sqrt{\frac{2}{\pi}} \frac{\Delta_{ {\Delta^\Pi}}}{\Delta_{th}} 
\exp \left( - \frac{ \Delta_{th}^2}{2 \Delta^2_{ {\Delta^\Pi}} } \right) .
\ee
This is an exponentially sensitive  function of the distance scale  $R$
over which we are averaging the perturbations.  

\begin{figure}[H]
		\includegraphics[width=0.95\textwidth]{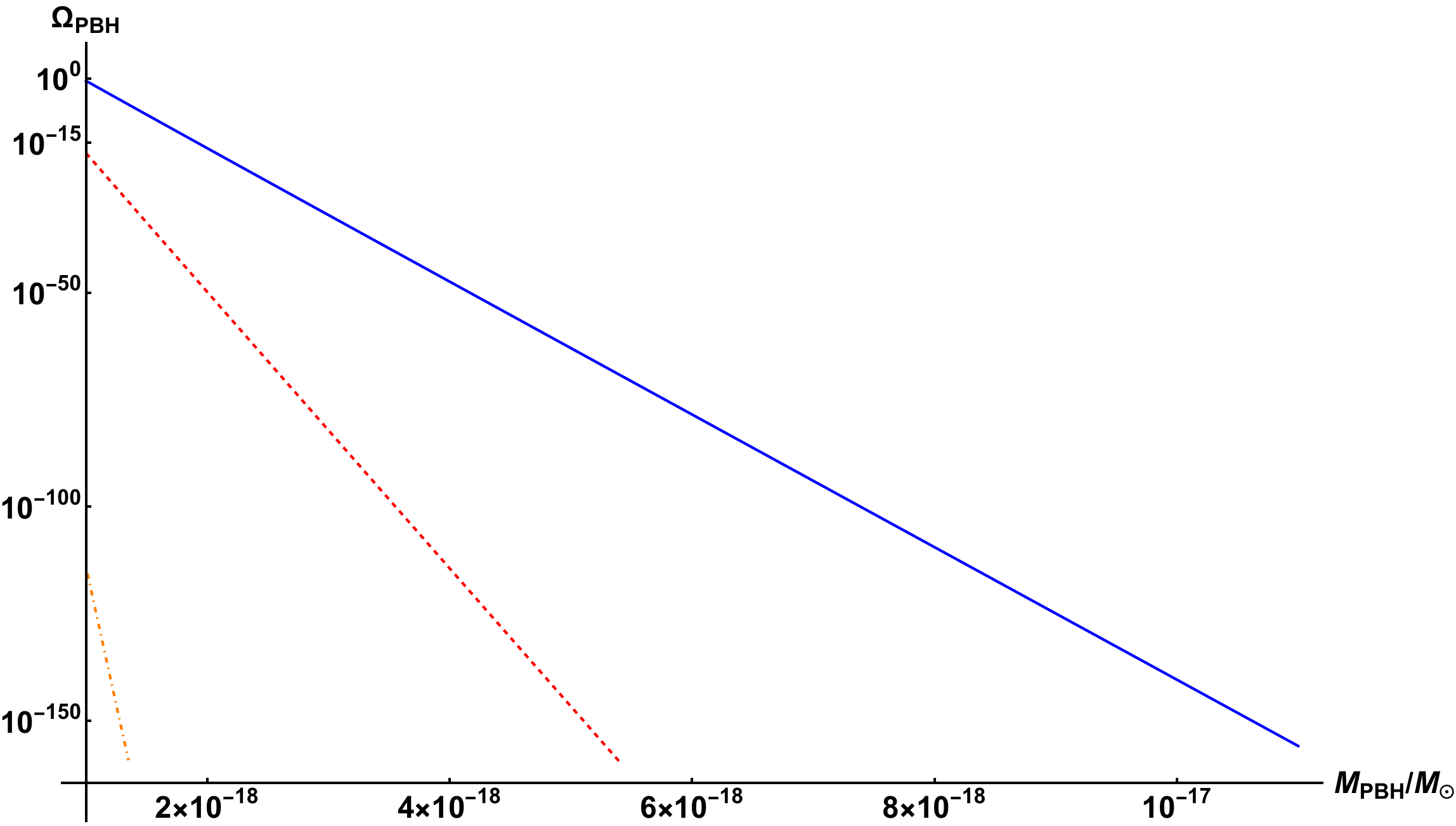}
	\caption{PBH abundance today, as a function of PBH mass/solar mass, calculated from eq. (\ref{abundanceToday}). Vertical axes in logarithmic scale. Solid blue corresponds to $c = 6.9297 \times 10^{-41}$, dashed red to $c = 10^{-40}$ and dot-dashed orange to $c = 2 \times 10^{-40} $.}
	\label{PBH_abundance}
\end{figure}

The PBH abundance today, $\Omega _{PBH} $ can be related to the abundance at the formation 
as follows \cite{Inomata:2017okj}:
\begin{align}\label{abundanceToday}
	\Omega_{PBH} (M_{PBH}) \simeq & 1.07 \times 10^{-44} \left( \frac{\alpha}{ 0.2 } \right)^{2} \left( \frac{106.75}{g_* (T_f)} \right)^{1/3}
	\left( \frac{\epsilon_{rad}}{c~\Delta_{th}} \right) \left( \frac{M_\odot}{M_{PBH}} \right) \\
	\nonumber {} & ~~~ \times \exp \left( - 1.46 \times 10^{99} \left( \frac{ 0.2 }{\alpha} \right) \left( \frac{g_* (T_f)}{106.75} \right)^{1/6} 
	\left( \frac{c~\Delta_{th}}{\epsilon_{rad}} \right)^2 \frac{M_{PBH}}{M_\odot} \right) .
\end{align}
 These calculations give the estimate for the abundances, which for a few different values of $c$ are plotted on Fig. \ref{PBH_abundance}. From this we see that, a small increase of PBH's mass decreases $\Omega_{PBH}$ by several orders in magnitude. As a consequence, only PBH's with the smallest masses, that have not decayed yet $(\sim 10^{-18} M_\odot)$, will contribute to $\Omega_{PBH}$.

For the values of the parameters $\alpha = 0.2, ~ g_* (T_f) = 106.75, ~ \Delta_{th} = 0.45, ~\epsilon_{rad} = 2, ~ M_{PBH} = 10^{-18} M_\odot$, and by demanding $\Omega_{PBH} < 0.26$, we constrain $c \geq 6.9297 \times 10^{-41} $. $\Omega_{PBH}$ is very sensitive to $c$, even a slight ($\mathcal{O} \left( 10^{-43} \right)$) change enhances/suppresses $\Omega_{PBH}$ by several orders in magnitude. 

Using the eq. (\ref{MpbhT}) we can rewrite the power spectrums, (\ref{PS_Delta_PI}) and (\ref{PS_Pi2}) as a function of PBH mass:
\bb\label{PS_num_vals}
\mathcal{P}_{\mathcal{R}^\Pi} \simeq
\frac{ 8.46 \times 10^{-103}}{c^2} \frac{M_\odot}{M_{PBH}} ;
\5\5\5
\mathcal{P}_{\Delta^\Pi} \simeq
\frac{ 3.38 \times 10^{-102} \epsilon_{rad}^2}{c^2} \frac{M_\odot}{M_{PBH}} .
\ee
For the PBH's of mass $=10^{-18} M_\odot$ and for $c = 6.9297 \times 10^{-41} $, we get $\mathcal{P}_{\mathcal{R}^\Pi} = 1.76 \times 10^{-4}$ (this corresponds to $q \simeq 5.89 \times 10^{-2}$ in eq. (\ref{PS_PBH})) and $\mathcal{P}_{\Delta^\Pi} = 2.82 \times 10^{-3}$. Note that the value of $\mathcal{P}_{\mathcal{R}^\Pi} $ is significantly larger than the one observed in CMB $\sim 10^{-9}$ which justifies our assumption (\ref{cond_1}).

It is worth pointing out that the lighter PBHs, with masses $<<10^{15}g$, are easier 
to produce in the present framework, since they require less tuning of the value of 
$c$. Such PBHs would have evaporated via the Hawking radiation before reaching present times. 
For instance, PBHs in the mass range $10^9g-10^{13}g$ would have evaporated during or
after BBN leading to modifications of the predictions for the light element abundances, 
and therefore their number density should be constrained, as it's reviewed in 
\cite{Carr:2020gox}. These and other interesting astrophysical 
constraints on the evaporating PBHs  were recently re-examined in \cite{Keith:2020jww} 
using novel software; all the constraints discussed in \cite{Keith:2020jww} apply to our case, too.

\vspace{0.1in}

We have considered above the PBH formation due to the large primordial fluctuations 
of the anomalyon generated during inflation. The anomalyon seems to be required 
by the consistency of the effective theory of gravity that accommodates the scale anomaly, and 
inflation remains to be the leading theory for the  production of 
the nearly scale-invariant spectrum consistent with the observations. 
However, the above mechanism assumes the existence  
of some new non-perturbative physics at high scale that sets the value of $a_g$.
Furthermore, the value of $c\sim 10^{-41}$ that gives rise to the abundance of PBH consistent 
with that of dark matter today, is unnaturally small. Such a small value would require 
an intricate model building around the high energy scale that fixes the value of $a_g$.
Given that $c=\sqrt{12}\lambda a_g$ with both $\lambda$ and $a_g$ being small, 
a natural value of it is $c\sim 10^{-30}$, for which the PBH abundance will be 
negligibly small as compared to $\Omega_{DM}\simeq 0.26$.

In the light of this, it might be interesting to comment on another 
anomalyon-mediated mechanism for the 
PBH formation that would not assume the new non-perturbative high energy phase. 
Ref. \cite{Amendola:2017xhl} proposed (see, \cite{Savastano:2019zpr}), 
and refs. \cite{Flores:2020drq,Domenech:2023afs} showed that if dark matter 
particles are attracted to one another 
by a long-range scalar force that is stronger than gravity, then these particles could collapse into a PBH during the radiation dominated epoch \cite{Amendola:2017xhl,Savastano:2019zpr,Flores:2020drq,Domenech:2023afs}\footnote
{The mechanism could in principle apply to ordinary particles too, 
assuming that they couple to a long-range scalar in 
the early universe, which acquires a mass later on to evade the fifth force constraints.}.
It is worth asking  a question whether the anomalyon  could play the role of the long-range scalar used in \cite{Amendola:2017xhl,Savastano:2019zpr,Flores:2020drq,Domenech:2023afs}.

Besides coupling to gravity, the anomalyon couples only to the trace anomalies, and if there are 
no gauge fields beyond the standard model, it will only couple to  the  gauge field kinetic terms of the $SU(3)\times SU(2)\times U(1)$ \cite{Riegert:1984kt}. These couplings are 
proportional  to $\sigma (\beta_j /\alpha_j) (FF)_j$, where $\s$ is the 
dimensionless anomalyon field, $\beta_j$ is the beta-function for the gauge 
coupling $\alpha_j$, and  $(FF)_j$ denotes  the gauge kinetic term for the respective gauge fields, $j=SU(3), SU(2), U(1)$.

Due to these interactions,  and due to the one-loop Feynman diagrams, the anomalyon will 
get effective couplings to the fermions charged under $SU(3)\times SU(2)\times U(1)$.
These induced couplings will be proportional to 
$\sigma \alpha_j   (\beta_j /\alpha_j)  m_k {\bar f}_k f_k$, where 
$f_k$ is a fermion of mass $m_k$ (i.e., a quark, electron, etc...).  If a dark matter particle is a massive 
fermion that has certain weak (or literally the weak) interactions with the standard model gauge fields, 
then the anomalyon will also have a similar coupling to the dark matter particle. 
This opens a door for the mechanism of \cite{Amendola:2017xhl,Savastano:2019zpr,Flores:2020drq,Domenech:2023afs} 
with the anomalyon providing the attraction between the dark matter particles.

Due to the interactions with the $SU(3)\times SU(2)\times U(1)$   gauge fields  at finite temperature of the radiation dominated universe the anomalyon will acquire a thermal mass, $m_A (T)$. The corresponding Compton wavelength, $1/m_A(T)$, should be larger than the short distance scale where the theory seizes to be valid. 
Furthermore, the anomalyon Compton wavelength 
should be much greater than the average inter-particle 
separation for the interactions due to the anomalyon 
to be considered long-range.  Last but not least,  the anomalyon necessarily has nonlinear 
self-interactions described by the Galileon terms.
This property distinguishes it from a scalar considered in \cite{Amendola:2017xhl,Savastano:2019zpr,Flores:2020drq,Domenech:2023afs}. 
As is well know, the Galileon interactions would lead to the Vainshtein 
screening of the scalar  force \cite {Vainshtein:1972sx,Deffayet:2001uk},  if the Vainshtein radius exceeds the average inter-particle separation. 

Thus, the above constraints need to be satisfied in order for the mechanism of  \cite{Amendola:2017xhl,Savastano:2019zpr,Flores:2020drq,Domenech:2023afs}
to be applicable in the present context. 
If the above constraints are satisfied, then the anamolyon exchange could provide a long-range force needed to start the collapse and PBH formation through the 
mechanism of \cite{Amendola:2017xhl,Savastano:2019zpr,Flores:2020drq,Domenech:2023afs}.  This possibility will require further study including a discussion of the effect of the Vainshtein screening on limiting growth.


\section{Tensor Perturbations}


The tensor perturbations (transverse and traceless) in (\ref{Lag_beg}), on FLRW background, are described by the following Lagrangian:
\begin{align}\label{tensor_perts}
	\nonumber \mathcal{L}_T = & \frac{a^2}{4} \Big( 1 - \frac{c^2}{12 a^2} - 2 \gamma^2 \frac{\mathcal{H}^2}{a^2} \Big) \Big( h_{i j}' \Big)^2
	- \frac{a^2}{4} \Big( 1 - \frac{c^2}{12 a^2} - 2 \gamma^2 \left( 1 - 2 \epsilon \right) \frac{\mathcal{H}^2}{a^2} \Big) 
	\Big( \partial_k h_{i j} \Big)^2 \\
	{} & + \frac{\mathfrak{c}}{2} \sigma_0 \Big[ \left( h_{ij}'' \right)^2 + \left( \partial_k^2 h_{ij} \right)^2 - 2 \left( \partial_k h_{ij}' \right)^2 \Big] - \frac{\mathfrak{c}}{2} \left( 1 - \epsilon \right) \mathcal{H}^2 \left( \partial_k h_{ij} \right)^2 .
\end{align}
The terms containing  $c^2$ are proportional to $(\M e^{-\s}/M)^2$, which 
is always much smaller than 1. Although these terms could be introducing small corrections to the conventional GR results, we will not keep track of them in what follows for simplicity.

It is convenient to work at the level of equations of motion. In addition, we take $\epsilon = const$. The equation of motion in this case reads
\begin{align}
	\nonumber {} & \left[ \mathfrak{c} \left( 1 - \epsilon \right) \mathcal{H}^2 - \frac{a^2}{2} \Big( 1 - 2 \gamma^2 \frac{\mathcal{H}^2}{a^2} \Big) \right]
	h_{i j}'' - a^2 \mathcal{H} \Big( 1 - 2 \gamma^2 \left( 1 - \epsilon \right) \frac{\mathcal{H}^2}{a^2} \Big) h_{i j}' \\
	{} & + \left[ \mathfrak{c} \left( 1 - \epsilon \right) \mathcal{H}^2 + \frac{a^2}{2} \Big( 1 - 2 \gamma^2 \left( 1 - 2 \epsilon \right) \frac{\mathcal{H}^2}{a^2} \Big) \right] \partial_k^2 h_{i j} \\
	\nonumber {} & + 2 \mathfrak{c} \mathcal{H} h_{i j}'''- 2 \mathfrak{c} \mathcal{H} \partial_k^2 h_{i j}'
	+ \mathfrak{c} \sigma_0 h_{i j}^{(IV)} + \mathfrak{c} \sigma_0 \partial_k^4 h_{i j}
	- 2 \mathfrak{c} \sigma_0 \partial_k^2 h_{i j}'' = 0 .
\end{align}
We expect the higher derivative terms to become relevant at energies comparable or above  the cut-off, and for that reason we discard them. The following field redefinition,
\bb\label{ansatz_diag}
h_{ij} = \frac{\sqrt{2}}{a} \left( 1 + \gamma^2 \frac{\mathcal{H}^2}{a^2}
+ \mathfrak{c} \frac{ (1 - \epsilon) }{\epsilon} \frac{\mathcal{H}^2}{a^2} \right) \mathbbm{h}_{ij}
+ \mathcal{O} \left( \gamma^4 ; \mathfrak{c}^2 ; \mathfrak{c} \gamma^2 \right) ,
\ee
eliminates the first time derivative and brings the equation of motion to the form:
\bb\label{reduced_EoM}
B_1 (\tau) \mathbbm{h}_{ij}'' + B_2 (\tau) \partial_k^2 \mathbbm{h}_{ij} + B_3 (\tau) \mathbbm{h}_{ij} = 0 ,
\ee
with the speed of propagation defined by
\bb\label{SoS_anom}
v_s^2 \equiv - \frac{B_2 (\tau)}{B_1 (\tau)} = 1 + 4 \gamma^2 H^2 \epsilon + 4 \mathfrak{c} H^2 \left( 1 - \epsilon \right) 
+ \mathcal{O} \left( \gamma^4 ; \mathfrak{c}^2 ; \mathfrak{c} \gamma^2 \right) ,
\ee
and 
\bb
\frac{B_3 (\tau)}{B_1 (\tau)} = a^2 H^2 \Big[ \epsilon - 2 - 6 \gamma^2 H^2 \epsilon
- 6 \mathfrak{c} H^2 \left( 1- 2 \epsilon \right) \Big]
+ \mathcal{O} \left( \epsilon^2 ; \gamma^4 ; \mathfrak{c}^2 ; \mathfrak{c} \gamma^2 \right) .
\ee
The equation (\ref{reduced_EoM}) can be obtained from the following Lagrangian:
\bb
\mathcal{L}_T = \frac12 \left( \mathbbm{h}_{ij}' \right)^2
- \frac12 v_s^2 \left( \partial_k \mathbbm{h}_{ij} \right)^2
- \frac12 a^2 H^2 \Big[ \epsilon - 2 - 6 \gamma^2 H^2 \epsilon 
- 6 \mathfrak{c} H^2 \left( 1 - 2 \epsilon \right) \Big] \mathbbm{h}_{ij}^2 .
\ee
First, let's rewrite the equation of motion in Fourier space:
\bb
\mathbbm{h}_k '' + \left[ v_s^2 k^2 - \frac{1}{\tau^2} \left( \nu^2 - \frac{1}{4} \right) \right] \mathbbm{h}_k = 0 \,,
\5\5 \text{with} \5\5
\nu = \frac{3}{2} + \epsilon + 2 \gamma^2 H^2 \epsilon + 2 \mathfrak{c} H^2 .
\ee
The following solution
\bb\label{tensor_pert_sol}
\mathbbm{h}_k = \frac{\sqrt{\pi}}{2} e^{i \frac{\pi}{2} \left( \nu + \frac12 \right)}
\sqrt{-\tau} H_\nu^{(1)} \left( - v_s k \tau \right) ,
\ee
satisfies the quantization criteria on sub-horizon scales $-k \tau \gg 1$:
\bb
\mathbbm{h}_k^* \mathbbm{h}_k' - \mathbbm{h}_k (\mathbbm{h}_k^*)' = - i .
\ee
The solution (\ref{tensor_pert_sol}) has the following asymptotic limits:
\bb
\mathbbm{h}_k (- k \tau \gg 1) = \frac{1}{\sqrt{2 v_s k}} e^{- i v_s k \tau} \,,
\5\5\5
| \mathbbm{h}_k (- k \tau \ll 1) | = \frac{1}{\sqrt{2 v_s k}} \left( \frac{v_s k}{ a H } \right)^{-\nu + \frac12} .
\ee
On super-horizon scales, $- k \tau \ll 1$, we also have:
\begin{align}
	\nonumber | h_k (- k \tau \ll 1) | = & \frac{\sqrt{2}}{a} \left( 1 + \gamma^2 H^2
	+ \mathfrak{c} \frac{ (1 - \epsilon) }{\epsilon} H^2 \right) | \mathbbm{h}_k (- k \tau \ll 1) | \\
	= & \left( 1 + \gamma^2 H^2 + \mathfrak{c} \frac{ (1 - \epsilon) }{\epsilon} H^2 \right) \frac{H}{(v_s k)^{3/2}} \left( \frac{v_s k}{ a H } \right)^{-\nu + \frac{3}{2}} .
\end{align}
The power spectrum of tensor fluctuations is
\begin{align}
	\nonumber \mathcal{P}_T = & \frac{k^3}{2 \pi^2} \Big( | {h}_{+} |^2 + | {h}_{\times} |^2 \Big) = \frac{4}{v_s^3} \left( 1 + 2 \gamma^2 H^2 + 2 \mathfrak{c} \frac{ (1 - \epsilon) }{\epsilon} H^2 \right)
	\left( \frac{H}{2 \pi} \right)^2 \left( \frac{v_s k}{ a H } \right)^{n_T} \\
	= & 4 \left( 1 + 2 \gamma^2 H^2 \left( 1 - 3 \epsilon \right) 
	+ 2 \mathfrak{c} H^2 \frac{ (1 - \epsilon) \left( 1 - 3 \epsilon \right) }{\epsilon}  \right)
	\left( \frac{H}{2 \pi} \right)^2 \left( \frac{k}{ a H } \right)^{n_T} ,
\end{align}
with
\bb
n_T = -2 \nu + 3 = - 2 \epsilon - \frac{4 \gamma^2 H^2 }{M^2} \epsilon - \frac{4 \mathfrak{c} H^2}{M^2}\,,
\ee
where  we restored $M$ for clarity. Then, the tensor to scalar ratio is
\bb
r = 16 \epsilon \left( 1 + \frac{ 2 \gamma^2 H^2 }{M^2} \left( 1 - 3 \epsilon \right) 
+ \frac{ 2 \mathfrak{c} H^2 }{M^2} \frac{ (1 - \epsilon) \left( 1 - 3 \epsilon \right) }{\epsilon}  \right) .
\ee
The ``consistency" relation coincides (to the leading order) with the 
one found in GR:
\bb
r = - 8 n_T .
\ee

Some comments are in order. From the formula (\ref{SoS_anom}), we see that unless $\mathfrak{c} < -\gamma^2 \epsilon/(1-\epsilon)$, the tensor fluctuations propagate with a speed greater than the speed of light. An interesting example is the one  with $\mathfrak{c} = 0$, in which case the equations of motion are of second order, and the tensor propagation speed is:
\bb
v_s \simeq 1 + \frac{2 \epsilon \gamma^2 H^2}{M^2} .
\ee 
Superluminality within an effective field theory is not necessarily a problem; such issues emerge even when one puts together  Quantum Electrodynamics with General relativity \cite{Drummond:1979pp}. To see how this is not a problem, one could compare  the propagation of a 
superluminal particle with a luminal one over a distance/time characteristic of the given space-time (for more details see \cite{Goon:2016une}). If the time/distance gained by the superluminal particle is less than the inverse of the EFT energy scale, then the effective field theory cannot be said to be problematic. 

In our case it is natural to choose the de Sitter horizon $(1/H)$ as the distance across which the comparison should takes place. Then, a superluminal mode with speed $v_s$, ``beats" the luminal one by the time given by 
\bb\label{time_diff}
\frac{1}{H} - \frac{1}{H v_s} = \frac{2 \epsilon \gamma^2 H}{M^2} + \mathcal{O} \left( \gamma^4 \right) .
\ee
If this time is shorter than $\bar{M}^{-1}$, then the superluminality can not be claimed within the EFT. Since $H<<M, \M<<M$ and $\epsilon<<1$ the condition 
\bb
\frac{2 \epsilon \gamma^2 H \bar{M} }{M^2} < 1 ,
\ee
can be satisfied for a huge range of the values of the parameter $\gamma$ 
without making any unnatural assumptions.


\section*{Acknowledgements}

We'd like to thank Zohar Komargodski, Alex Kusenko, Shinji Mukohyama, Misha Shifman, and Paul Steinhardt  for valuable communications. 
GG is supported in part by the NSF grant PHY-2210349. 
GT is supported by the Simons Foundation grant number 654561.


\appendix


\section{Equations of Motion}\label{app_EoM}

The variation of the Lagrangian (\ref{Lag_beg}), w.r.t. the metric and $\sigma$ generates equations (\ref{metric_var}) and (\ref{sigma_var}) respectively. The notations are as follows:
\bb
\mathcal{J}_{\mu \nu} \equiv G_{\mu \nu} + 2 \Sigma_{\mu \nu} - 2 g_{\mu \nu} \left[ \Sigma \right] + 2 \nabla_\mu \sigma \nabla_\nu \sigma 
+ g_{\mu \nu} \left( \partial \sigma \right)^2 \,,
\ee
\begin{align}
	\nonumber \mathcal{K}_{\mu \nu} \equiv & \Big( W_{\mu \alpha \nu \beta} - \frac12 g_{\mu \nu} R_{\alpha \beta} - \frac12 g_{\alpha \beta} R_{\mu \nu}
	+ \frac12 g_{\mu \beta} R_{\nu \alpha} + \frac12 g_{\nu \alpha} R_{\mu \beta} \\ 
	\nonumber {} & ~~ + \frac{1}{3} g_{\mu \nu} g_{\alpha \beta} R
	- \frac{1}{3} g_{\mu \beta} g_{\nu \alpha} R \Big) \Sigma^{\alpha \beta}  + \Sigma_{\mu \nu}^2 - \left[ \Sigma \right] \Sigma_{\mu \nu}
	+ \frac12 g_{\mu \nu} \Big( \left[ \Sigma \right]^2 - \left[ \Sigma^2 \right] \Big) \\ 
	{} & + \Big( W_{\mu \alpha \nu \beta} - \frac12 g_{\mu \nu} R_{\alpha \beta} + \frac12 g_{\mu \beta} R_{\nu \alpha}
	+ \frac12 g_{\nu \alpha} R_{\mu \beta} + \frac{1}{12} g_{\mu \nu} g_{\alpha \beta} R \\
	\nonumber {} & ~~~~~~ - \frac{1}{3} g_{\mu \beta} g_{\nu \alpha} R \Big) \partial^\alpha \sigma \partial^\beta \sigma + \partial_\mu \sigma \partial_\alpha \sigma \Sigma^\alpha_\nu + \partial_\nu \sigma \partial_\alpha \sigma \Sigma^\alpha_\mu - \partial_\mu \sigma \partial_\nu \sigma \left[ \Sigma \right] \\
	\nonumber {} & + \partial_\mu \sigma \partial_\nu \sigma \left( \partial \sigma \right)^2
	- g_{\mu \nu} \left( \partial_\alpha \sigma \partial_\beta \sigma \Sigma^{\alpha \beta}
	+ \frac{1}{4} \left( \partial \sigma \right)^4 \right) \,,
\end{align}
\bb
T_{\mu \nu} \equiv \partial_\mu \omega \partial_\nu \omega -  g_{\mu \nu} \Big( \frac12 \left( \partial \omega \right)^2 + V (\omega) \Big) \,,
\ee
\bb
\mathcal{C}_{\mu \nu} \equiv \frac{1}{\sqrt{g}} \frac{\delta}{\delta g^{\mu \nu}} \Big( \sigma \sqrt{g} W_{\alpha \beta \epsilon \rho} W^{\alpha \beta \epsilon \rho} \Big) \,,
\ee
\bb
\Sigma_{\mu \nu} \equiv \nabla_\mu \nabla_\nu \sigma \,,
\5\5\5\5\5
\left[ X \right] \equiv g^{\mu \nu} X_{\mu \nu} .
\ee
If one assumes the background metric is given by the spatially flat FLRW (\ref{FRW_ansatz}), then $tt$, $xx$ part of the metric variation, sigma variation, and their combination -- the anomalous equation (\ref{cov_anom_eq}), become respectively:
\bb\label{set_1}
H^2 - \lambda^2 e^{-2 \sigma} \left( \dot{\sigma} - H \right)^2  + \frac{\gamma^2}{M^2}
\left[ \left( \dot{\sigma} - H \right)^4 - H^4 \right] = \frac{1}{6 M^2} T_{tt} \,,
\ee
\begin{align}\label{set_2}
	\nonumber {} & 2 M^2 \dot{H} - 4 \gamma^2 H^2 \dot{H} \\
	{} & + 2 \Big( M^2 \lambda^2 e^{-2 \sigma} - 2 \gamma^2 \left( \dot{\sigma} - H \right)^2 \Big) \Big( \partial_t \left( \dot{\sigma} - H \right) + H \left( \dot{\sigma} - H \right) + \left( \dot{\sigma} - H \right)^2 \Big) \\
	\nonumber {} & + 3 M^2 H^2 - 3 M^2 \lambda^2 e^{-2 \sigma} \Big( \dot{\sigma} - H \Big)^2
	+ 3 \gamma^2 \left[ \Big( \dot{\sigma} - H \Big)^4 - H^4 \right] = - \frac12 T_x^x \,,
\end{align}
\bb\label{set_3}
\lambda^2 e^{-2 \sigma}
\left[ \left( \dot{\sigma} - H \right)^2 - H \left( \dot{\sigma} - H \right) - \partial_t \left( \dot{\sigma} - H \right)  \right]
+ \frac{2 \gamma^2}{3 M^2 a^3} ~\partial_t \left[ a^3 \left( \dot{\sigma} - H \right)^3 \right] = 0 \,,
\ee
\bb\label{set_4}
\dot{H} + 2 H^2 - 2 \frac{\gamma^2}{M^2} H^2 \Big( \dot{H} + H^2 \Big) = \frac{1}{12 M^2} \Big( T_{tt} - 3 T_x^x  \Big) \,.
\ee
These equations must be supplemented by the equation of the canonical scalar field, which is convenient to write as the continuity equation (\ref{cont_eq}). Not all of the above four equations are independent. One can show that by using (\ref{set_1}) with  the continuity equation (\ref{cont_eq}), and with any of (\ref{set_2}), (\ref{set_3}) or (\ref{set_4}), one can recover the remaining two equations. As discussed in section \ref{sect_1}, it is convenient to choose (\ref{set_1}), (\ref{set_4}) and (\ref{cont_eq}) as the main set describing the dynamics.


\section{The Strong Coupling Scale}


Tree level scattering amplitudes of excitations described by the  trace-anomaly effective action 
(\ref {actiontotal}) exhibit strong coupling in Minkowski space-time at a certain high-energy scale $\M$. 
In a curved background, e.g, in the cosmological FLRW spacetime, the strong coupling  
scale may differ from $\M$. 
The latter can be calculated 
by comparing the coefficients in front  of the quadratic and  nonlinear terms of the perturbations 
in a respective FLRW background.  Before that, however, one needs to establish  the EFT restrictions 
on the FLRW background itself, i.e., restrictions on the values of 
the spacetime curvature,  and on the derivatives of the $\s$ field constituting the 
cosmological background. 

As noted in Section \ref{into_sum},  the class of FLRW cosmological solutions discussed in this work
has  the metric $\g_{\mu\nu}=e^{-2\s}g_{\mu\nu}$ proportional to  the Minkowski spacetime metric $\eta_{\mu\nu}$.
As a consequence, all curvature invariants of  $\g$ in (\ref {actiontotal}) and (\ref {GRbar}) vanish.  
Therefore, the EFT restrictions imposed on the value of the curvatures in (\ref {actiontotal}) coincide with
the known restrictions imposed by the GR action (\ref {GR}). Hence, we conclude that for the specific class 
of the FLRW solutions described above, the EFT constrains 
the characteristic  mass scale of the curvatures of the metric 
$g$  to be smaller than $M\sim M_{Pl}$.  On the other hand, there are two 
terms in (\ref{actiontotal}) that contain no curvatures of $\g$, but depend on 
derivatives of $\s$, as in the cubic and quartic Galileons. The strong scale of these terms is $\M$.  Because there are only two such therms (as opposed to an infinite number of them)  
they can be treated exactly. Indeed, they have been retained in our equations 
as exact terms without truncation. This treatment enables classical background solutions with the 
derivatives of $\s$ exceeding $\M$ to be consistently included.

Let us now turn to the perturbations about the FLRW background. 
We write the  coefficient in front of the quadratic term for $\sigma$ field 
as follows:
\bb
-6\sqrt{g} g^{00} \M^2 e^{-2(\s_I+ \s_c(\tau))}\,,
\label{strong0}
\ee
where $g^{00}$ is a component of the inverse metric, 
$\sigma_I$ is an integration constant denoting the initial value of the classical solution for $\s$
\bb
\s_{0}=\s_I +\s_c = ln ( {\sqrt {12} \lambda a_{I} / c}  ) + ln ( {a(t)/ a_I} )\,,
\label{decompose}
\ee
$a_I$ is an arbitrary initial value of the scale factor, 
while $\sigma_c(\tau)$ is a classical background solution  with the 
initial value equal to zero.

Let us focus on a generic FLRW solution with $g_{\mu\nu} = a^2(\tau)\eta_{\mu\nu}$. 
The  coefficient in front of the quadratic term for $\s_c$ is 
\bb
-6\sqrt{g} g^{00} \M^2 e^{-2\s_0}\,=  6\M^2 e^{-2\s_I}\,a^2 \,e^{-2\s_c}\,=6\M^2 e^{-2\s_I}\,a_I^2 \,.
\label{strong00}
\ee
This coefficient is time independent due to the cancellation of time dependence between 
the scale factor  and the background solution for $\s$; moreover, the value of this 
coefficient, $6\M^2 e^{-2\s_I}\,a_I^2 = c^2 M_{Pl}^2/2$, depends only on one integration constant 
$c$.

We proceed further and look at the perturbations that can be parametrized as follows
\bb
\s = \s_I + \s_c(\tau) + \delta \s (\tau, {x})\,.
\label{pert}
\ee
Substituting the above into the action we get  the quadratic and leading nonlinear terms
in the following form:
\bb
\sqrt{g_{FLRW}} \left ( -{12\over 2}\M^2 e^{-2(\s_I+\s_c)}  \, (\nabla \delta\s)^2 + {\tilde \gamma}_3 \,\nabla^2 \delta \s (\nabla \delta\s)^2
+ {\tilde \gamma}_4 (\nabla \delta\s)^2(\nabla \delta\s)^2\right   ) \cdots,
\label{deltasigma_c}
\ee
where the covariant derivatives are w.r.t. the background FLRW metric, and 
${\tilde \gamma}_{3,4}$ are dimensionless coefficients determined by the particle content 
of a QFT that gives rise to the trace anomaly; it is sufficient for our purposes  
to approximate ${\tilde \gamma}_{3,4}\sim 1$.

We've already shown in (\ref {strong00}) that the coefficient of the quadratic term is time independent;
remarkably, the coefficients of both leading nonlinear terms in (\ref {deltasigma_c})  are also 
time independent because $\sqrt{g} g^{00}g^{00}|_{FLRW} =1$; this is a consequence of the 
cubic and quartic Galileons being invariant under conformal rescaling of metric. 
Taking these considerations into account, one can easily deduce from 
(\ref {deltasigma_c}) that the strong scale of the co-moving momenta and energy
of perturbations  $\delta \sigma_c$  is:
\bb
\Lambda_c \equiv \sqrt{12}\, \M e^{-\s_I}a_I = c \,M_{Pl} \,=\sqrt{12}\,a_g\,\M\,. 
\label{strong}
\ee
The value of $\Lambda_c$ depends on the choice of the initial value of the anomalyon
$\sigma_I$,  and of the scale factor $a_I$, which combine into the dependence on 
one constant  $c = \sqrt{12}\,a_g\,\lambda$  
(the same co-moving cutoff is obtained from the anomalyon perturbations generated 
by the ``$\cdots$ terms" in (\ref{GRbar})).

Furthermore,  we should emphasize that the  value of $\M e^{-\s}$ should never  exceed $M$ in order for 
the theory to have a positive effective Planck mass squared. This imposes a constraint 
\bb
{\M} e^{-\s_I} a_I < M a\,,
\label{Mpl_const}
\ee
and since $a_I\leq a$, then it is sufficient to satisfy $e^{-\s_I} <M/\M$.   
As an example,  if $\M \sim 10^8 GeV$, then $\s_I$ could be $\s_I \geq -20$,  and 
the latter condition readily satisfied.  At the same time, the value of 
$\s_0=\s_I+\s_c$ would change from $\s_0 \simeq -20$, at  the beginning of inflation, to  
$\s_0 \simeq 40$ by the end of 60 e-folds, with the value of $\Sigma=\M e^{-\s}$ 
ranging  from $10^{16} GeV$ during inflation  
and decreasing to $0.1eV$ by the end of inflation. Note that 
after the QCD phase transition (or QCD-like high-energy phase transition if such is present) 
the anomalyon acquires a potential with a minimum near $\s=0$. 
At that point the anomalyon acquires a mass, and depending on the parameters could itself act as a 
dark matter particle. This particle would couple to  $({\vec E}^2 -{\vec B}^2)$
of the electromagnetic field  \cite {followUpPaper}.


\bibliographystyle{utphys}
\bibliography{refs}


\end{document}